\documentclass[aps,prd,notitlepage,nofootinbib,superscriptaddress,twocolumn]{revtex4-2}
\usepackage[utf8]{inputenc}

\usepackage[table ]{ xcolor}

\usepackage{multirow}

\usepackage{hyperref}
\usepackage{graphicx}
\usepackage{mathtools}
\usepackage{amsmath,amssymb,amsfonts,amsthm,latexsym,stmaryrd,physics,mathrsfs}
\allowdisplaybreaks[4]
\usepackage{marginnote}
\usepackage{color}
\usepackage{bm}
\usepackage{float}
\usepackage{nicefrac}
\usepackage{microtype} 
\usepackage{booktabs}
\usepackage{verbatim}
\usepackage{setspace}
\usepackage[T1]{fontenc}
\usepackage[utf8]{inputenc}
\usepackage{stfloats}
\usepackage{diagbox}
\usepackage{dsfont}

\hypersetup{
	colorlinks=true
,urlcolor=blue
,anchorcolor=blue
,citecolor=blue
,filecolor=blue
,linkcolor=blue
,menucolor=blue
,pagecolor=blue
,linktocpage=true
,pdfproducer=medialab
}

\def\beq{\begin{equation}}
\def\eeq{\end{equation}}
\newcommand{\bea}{\begin{eqnarray}}
\newcommand{\eea}{\end{eqnarray}}
\def\bi{\begin{itemize}}
\def\ei{\end{itemize}}
\def\ba{\begin{array}}
\def\ea{\end{array}}
\def\bfig{\begin{figure}}
\def\efig{\end{figure}}

\def\C{\mathbb{C}}
\def\R{\mathbb{R}}
\def\Z{\mathbb{Z}}

\newcommand{\Slc}{\mathrm{SL}(2,\mathbb{C})}

\newcommand{\Su}{\mathrm{SU}(2)}

\def\be{\begin{eqnarray}}
\def\ee{\end{eqnarray}}

\newcommand{\ci}{\mathcal I}
\newcommand{\cj}{\mathcal J}
\newcommand{\ck}{\mathcal K}

\newcommand{\cs}{\mathcal S}

\newcommand{\cz}{\mathcal Z}

\newcommand{\fa}{\mathfrak{a}}

\newcommand{\fj}{\mathfrak{j}}

\newcommand{\fn}{\mathfrak{n}}

\renewcommand{\a}{\alpha}

\newcommand{\g}{\gamma}

\newcommand{\sig}{\sigma}

\renewcommand{\l}{\lambda}

\renewcommand{\O}{\Omega}
\renewcommand{\t}{\tau}

\newcommand{\rmd}{\mathrm d}

\newcommand{\lt}{\left}
\newcommand{\rt}{\right}

\newcommand{\act}{\rhd}

\newcommand{\sn}{\mathscr{N}}

\newcommand{\re}{\mathrm{Re}}

\begin{document}

\title{Complex critical points and curved geometries in four-dimensional Lorentzian spinfoam quantum gravity}

\author{Muxin Han}
\email{hanm(AT)fau.edu}
\affiliation{Department of Physics, Florida Atlantic University, 777 Glades Road, Boca Raton, FL 33431-0991, USA}
\affiliation{Institut f\"ur Quantengravitation, Universit\"at Erlangen-N\"urnberg, Staudtstr. 7/B2, 91058 Erlangen, Germany}

\author{Zichang Huang}
\email{hzc881126(AT)hotmail.com}
\affiliation{Department of Physics, Center for Field Theory and Particle Physics,
	and Institute for Nano- electronic devices and Quantum computing, Fudan University, Shanghai 200433, China}
\affiliation{State Key Laboratory of Surface Physics, Fudan University, Shanghai 200433, China}

\author{Hongguang Liu} 
\email{hongguang.liu(AT)gravity.fau.de}
\affiliation{Institut f\"ur Quantengravitation, Universit\"at Erlangen-N\"urnberg, Staudtstr. 7/B2, 91058 Erlangen, Germany}

\author{Dongxue Qu}
\email{dqu2017(AT)fau.edu}
\affiliation{Department of Physics, Florida Atlantic University, 777 Glades Road, Boca Raton, FL 33431-0991, USA}

\begin{abstract}

This paper focuses on the semiclassical behavior of the spinfoam quantum gravity in 4 dimensions. There has been long-standing confusion, known as the flatness problem, about whether the curved geometry exists in the semiclassical regime of the spinfoam amplitude. The confusion is resolved by the present work. By numerical computations, we explicitly find curved Regge geometries from the large-$j$ Lorentzian Engle-Pereira-Rovelli-Livine (EPRL) spinfoam amplitudes on triangulations. These curved geometries are with small deficit angles and relate to the complex critical points of the amplitude. The dominant contribution from the curved geometry to the spinfoam amplitude is proportional to $e^{i \ci}$, where $\ci$ is the Regge action of the geometry plus corrections of higher order in curvature. As a result, the spinfoam amplitude reduces to an integral over Regge geometries weighted by $e^{i \ci}$ in the semiclassical regime. As a byproduct, our result also provides a mechanism to relax the cosine problem in the spinfoam model. Our results provide important evidence supporting the semiclassical consistency of the spinfoam quantum gravity.

\end{abstract}

\maketitle



The semiclassical consistency is an important requirement in quantum physics. Any satisfactory quantum theory must reproduce the corresponding classical theory in the approximation of small $\hbar$. In particular, the role of semiclassical analysis is more crucial in the field of quantum gravity. Due to the limitation of experimental tests, the semiclassical consistency is one of only few physical constraints for quantum gravity: a satisfactory quantum theory of gravity must reproduce General Relativity (GR) in the semiclassical regime.

This paper focuses on the semiclassical analysis of Loop Quantum Gravity (LQG). LQG as a background-independent and non-perturbative approach has been demonstrated to be a competitive candidate toward the final quantum gravity theory (see e.g., \cite{Thiemann:2007pyv,Rovelli:2014ssa,Perez2012,Rovelli:2010bf,Ashtekar:2017yom,Ashtekar:2021kfp} for reviews). The path integral formulation of LQG, known as the \emph{spinfoam theory} \cite{Engle:2007uq,Engle:2007wy,Rovelli:2005yj,Livine:2007vk,Freidel:2007py}, is particularly interesting for testing the semiclassical consistency of LQG, because of the connection between the semiclassical approximation of path integral and the stationary phase approximation. A central object in the spinfoam theory is the \emph{spinfoam amplitude}, which defines the covariant transition amplitude of LQG. The recent semiclassical analysis reveals the interesting relation between spinfoam amplitudes and the Regge calculus, which discretizes GR on triangulations \cite{Conrady:2008mk,Barrett:2009mw,Han:2011re,Han:2013gna,Kaminski:2017eew,Liu:2018gfc,Simao:2021qno,Dona:2020yao}. This relation makes the semiclassical consistency of the spinfoam theory promising.

Nevertheless, it has been argued that an accidental flatness constraint might emerge in the semiclassical regime, so that spinfoam amplitudes would be dominated by only flat Regge geometries, whereas curved geometries were absent \cite{Engle:2020ffj,Hellmann:2012kz,Bonzom:2009hw,Han:2013hna,Gozzini:2021kbt}. The suspicion of lacking curved geometry in the semiclassical regime has lead to the doubt about the semiclassical behavior. This \emph{flatness problem} has been a key issue in the spinfoam LQG for more than a decade.  

In this work, we resolve the flatness problem by explicitly finding curved Regge geometries from the 4-dimensional Lorentzian EPRL spinfoam amplitude. These curved geometries are with small deficit angles $\delta_h$, and have been overlooked in the model because they correspond to complex critical points slightly away from the real integration domain. But they can be revealed by a more refined stationary phase analysis involving the analytic continuation of the spinfoam integrand. These curved Regge geometries still give non-suppressed dominant contributions to the spinfoam amplitude. The contributions are proportional to $e^{i \ci}$ where $\ci$ is the Regge action of the curved geometry plus corrections of the second and higher orders in $\delta_h$. The spinfoam amplitude reduces to an integral over Regge geometries weighted by $e^{i \ci}$ in the semiclassical regime.

\begin{figure}[h]	
	\centering
	\includegraphics[scale=0.25]{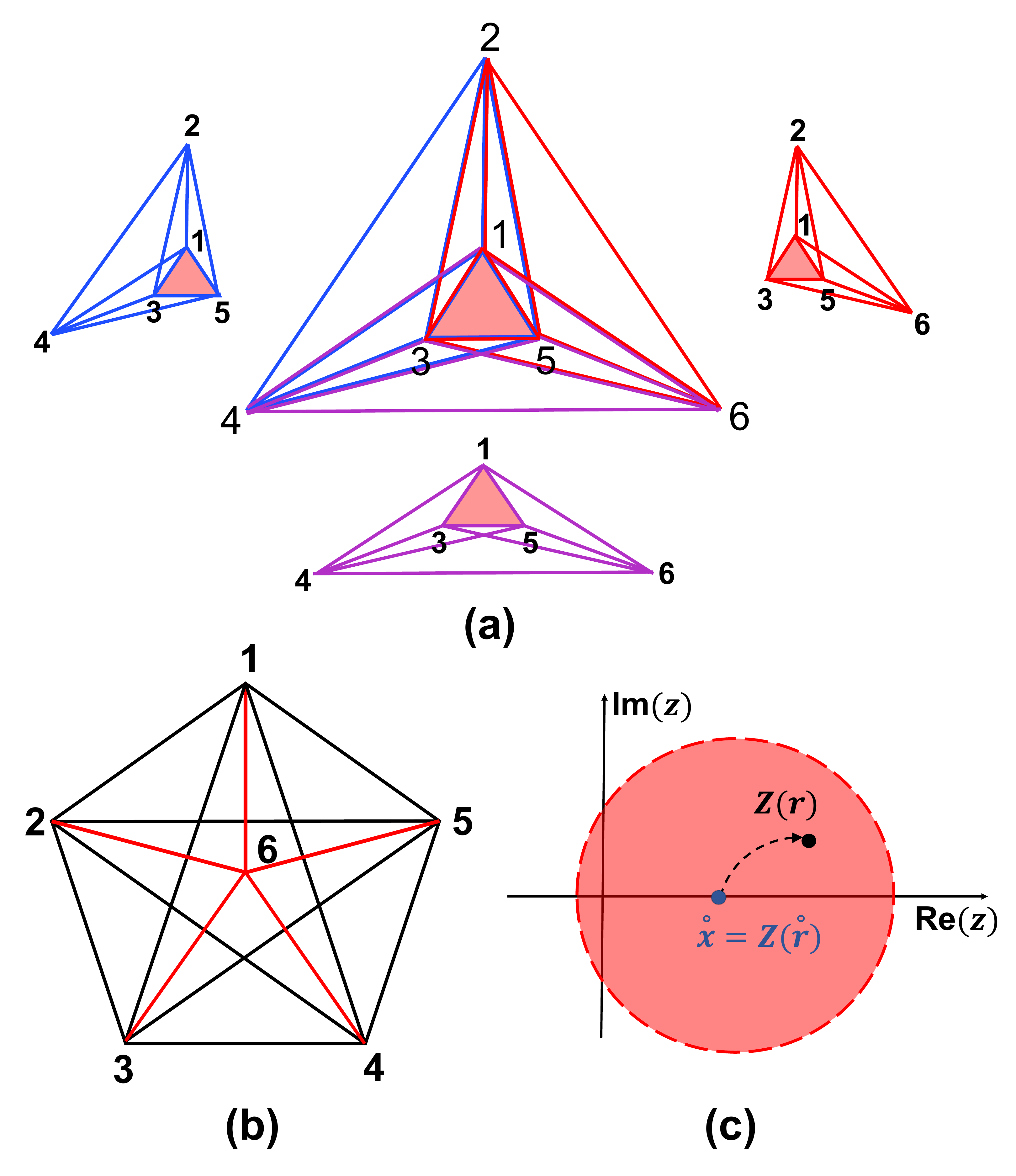}
	\caption{(a) The $\Delta_3$ triangulation (the center panel) made by gluing three 4-simplices (in blue, red, and purple). The internal triangle $(135)$ is highlighted in red. (b) The triangulation $\sig_{\text{1-5}}$ made by the 1-5 Pachner move dividing a 4-simplex into five 4-simplices. $\sig_{\text{1-5}}$ has 10 internal triangles and 5 internal segments $I=1,\cdots,5$ (red). (c) The real and complex critical points $\mathring{x}$ and $Z(r)$. $\cs(r,z)$ is analytic extended from the real axis to the complex neighborhood illustrated by the red disk.}
	\label{pic}
\end{figure}

These results are illustrated by the numerical analysis of the EPRL spinfoam amplitudes on triangulations $\Delta_3$ and $\sig_{\text{1-5}}$ (FIG.\ref{pic}(a) and (b)). As a byproduct, the ``cosine problem'' \cite{Engle:2011pmf} is shown to be relaxed on $\Delta_3$.

Our results provide important evidence supporting the semiclassical consistency of the spinfoam theory.




\emph{Spinfoam amplitude.}---The 4-dimensional triangulation $\ck$ contains 4-simplices $v$, tetrahedra $e$, triangles $f$, line segments, and points. We denote the internal triangle by $h$ and the boundary triangle by $b$ ($f$ is either $h$ or $b$), and assign the SU(2) spins $j_h,j_b\in\mathbb{N}_0/2$ to internal and boundary triangles $h,b$. The spin $j_f=j_h$ or $j_b$ relates to the quantum area of $f$ by $\fa_f=8\pi\g G\hbar\sqrt{j_f(j_f+1)}$ \cite{Rovelli1995,ALarea}. The Lorentzian EPRL spinfoam amplitude on $\ck$ sums over internal spins $\{j_h\}$:
\be
A(\ck)&=&\sum_{\{j_{h}\}}\prod_h \bm{d}_{j_h}\int [\rmd g\rmd\mathbf{z}]\, e^{S
\left(j_{h}, g_{v e}, \mathbf{z}_{vf};j_b,\xi_{eb}\right)}, \label{amplitude}\\
&& [\rmd g\rmd \mathbf{z}]=\prod_{(v, e)} \mathrm{d} g_{v e} \prod_{(v,f)} \mathrm{d}\O_{\mathbf{z}_{v f}},
\ee
where $\bm{d}_{ j_h}=2j_h+1$. The spinfoam action $S$ is complex and linear to $j_h,j_b$ \cite{Han:2013gna}. The boundary states of $A(\ck)$ are SU(2) coherent states $|j_b,\xi_{eb}\rangle$ where $\xi_{eb}=u_{eb}\act(1,0)^T$, $u_{eb}\in \Su$. $j_b,\xi_{eb}$ determines the area and the 3-normal of $b$ in the boundary tetrahedron $e$. The summed/integrated variables are $g_{ve}\in\Slc$, $\mathbf{z}_{vf}\in\mathbb{CP}^1$, and $j_h$. The boundary $j_b,\xi_{eb}$ are not summed/integrated. $\rmd g_{ve}$ is the $\Slc$ Haar measure. $\mathrm{d}\O_{\mathbf{z}_{v f}}$ is a scaling invariant measure on $\mathbb{CP}^1$. A cut-off $j_h^{\rm max}$ should be imposed if $\sum_{j_h}$ leads to divergence.

By the area spectrum, the classical area $\fa_f$ and small $\hbar$ imply the large spin $j_f\gg1$. This motivates to understand the large-$j$ regime as the semiclassical regime of $A(\ck)$. To probe the semiclassical regime, we scale uniformly $\{j_b,j_h\}\rightarrow\{\lambda j_b,\lambda j_h\}$, where $\lambda\gg1$. Scaling spins implies $S\rightarrow\lambda S$. Moreover, it is convenient to apply the Poisson summation formula to replace the sum over $j_h$ by integral 
\be
\!\!\!\!\!\!A(\ck)&=&\sum_{\{k_h\in\mathbb{Z}\}} \int \prod_h\mathrm{d} j_{h}\prod_h\lt(2\l \bm{d}_{\l j_{h}}\rt)\int [\rmd g\rmd \mathbf{z}] \,e^{\lambda S^{(k)}},\label{integralFormAmp}\\
&&S^{(k)}=S+4\pi i \sum_h j_h k_h, 
\ee
where $j_h$ is real and continuous.The details of $A(\ck)$, $S$, and Poission summation are reviewed in Appendix \ref{Spinfoam amplitude, Poisson resummation formula, and parametrization}.

\emph{Real critical points and flatness.}---For each ${k_h}$ in \eqref{integralFormAmp}, by the stationary phase method, the integral with $\l\gg1$ is approximated by the dominant contributions from solutions of the critical equations 
\be
\re(S)&=&\partial_{g_{ve}}S=\partial_{\mathbf{z}_{vf}}S=0,\label{eom1}\\
\partial_{j_h}S&=&4\pi i k_h, \qquad k_h\in\Z.\label{eom2}
\ee
The solution inside the integration domain is denoted by $\{\mathring{j}_h,\mathring{g}_{ve},\mathring{\bf z}_{vf}\}$. We view the integration domain as a real manifold, and call $\{\mathring{j}_h,\mathring{g}_{ve},\mathring{\bf z}_{vf}\}$ the \textit{real critical point}. 

Every solution satisfying the part \eqref{eom1} and a nondegeneracy condition endows a Regge geometry to $\ck$ with 4d orientation \cite{Han:2013gna,Han:2011re,Barrett:2009mw,Conrady:2008mk}. Further imposing \eqref{eom2} to these Regge geometries gives the accidental flatness constraint to every deficit angle $\delta_h$ hinged by the internal triangle $h$ \cite{Bonzom:2009hw,Han:2013hna}
\be
\g \delta_h= 4\pi k_h, \qquad k_h\in\Z. \label{flat}
\ee
The Barbero-Immirzi parameter $\g\neq 0$ is finite. When $k_h=0$, $\delta_h$ at every internal triangle is zero, so the Regge geometry endowed by the real critical point is flat. If the dominant contribution to $A(\ck)$ with $\l\gg1$ only comes from real critical points, Eq.\eqref{flat} implies that only the flat geometry and
geometries with $\g\delta_h=\pm 4\pi\mathbb{Z}_+$ can contribute dominantly to $A(\ck)$, whereas the contributions from generic curved geometries are suppressed. If this was true, the semiclassical behavior of $A(\ck)$ would fail to be consistent with GR.

A generic $\{\mathring{j}_h,\mathring{g}_{ve},\mathring{\bf z}_{vf}\}$ can endow discontinuous 4d orientation, i.e., the orientation flips between 4-simplices. Then \eqref{flat} becomes $\g \sum_{v\in h}s_v\Theta_{h}(v)= 4\pi k_h$ where $s_v=\pm1$ labels two possible orientations at each $4$-simplex $v$. $\Theta_{h}(v)$ is the dihedral angle hinged by $h$ in $v$.

\emph{Complex critical points.}---As we will show, the large-$\l$ spinfoam amplitude does receive non-suppressed contributions from curved geometries with small but nonzero $|\delta_h|$. Demonstrating this property needs a more refined stationary phase analysis for the complex action with parameters \cite{10.1007/BFb0074195,Hormander}: We consider the large-$\l$ integral $\int_K e^{\l S(r,x)}\rmd^N x$, and regard $r$ as parameters. $S(r,x)$ is an analytic function of $r\in U\subset \R^k,x\in K\subset \R^N$. $U\times K$ is a neighborhood of $(\mathring{r},\mathring{x})$. $\mathring{x}$ is a real critical point of $S(\mathring{r},x)$. $\mathcal{S}(r,z)$, $z=x+i y \in \mathbb{C}^{N}$, is the analytic extension of $S(r,x)$ to a complex neighborhood of $\mathring{x}$. The complex critical equation $\partial_{z} \mathcal{S}=0$ is solved by ${z}=Z(r)$ where $Z(r)$ is an analytic function of $r$ in the neighborhood $U$. When $r=\mathring{r}$, $Z(\mathring{r})=\mathring{x}$ reduces to the real critical point. When $r$ deviates away from $\mathring{r}$, $Z(r)\in\C^N$ can move away from the real plane $\R^N$, thus is called the \emph{complex critical point} (see FIG.\ref{pic}(b)). We have the following large-$\l$ asymptotic expansion for the integral
	\be
	\int_K e^{\lambda S(r,x)}  \mathrm{d}^N x &=& \left(\frac{1}{\lambda}\right)^{\frac{N}{2}} \frac{e^{\lambda \mathcal{S}(r,Z(r))}}{\sqrt{\det\left(-\delta^2_{z,z}\mathcal{S}(r,Z(r))/2\pi\right)}}\nonumber\\ 
	&&\times \lt[1+O(1/\l)\rt]
	\label{asymptotics}
	\ee
where $\cs(r,Z(r))$ and $\delta^2_{z,z}\mathcal{S}(r,Z(r))$ are the action and Hessian at the complex critical point.

The crucial information of \eqref{asymptotics} is: the integral can receive the dominant contribution from the complex critical point away from the real plane. This fact has been overlooked by the argument of the flatness problem.


\begin{figure}[h]
	\centering
	\includegraphics[width=0.5\textwidth]{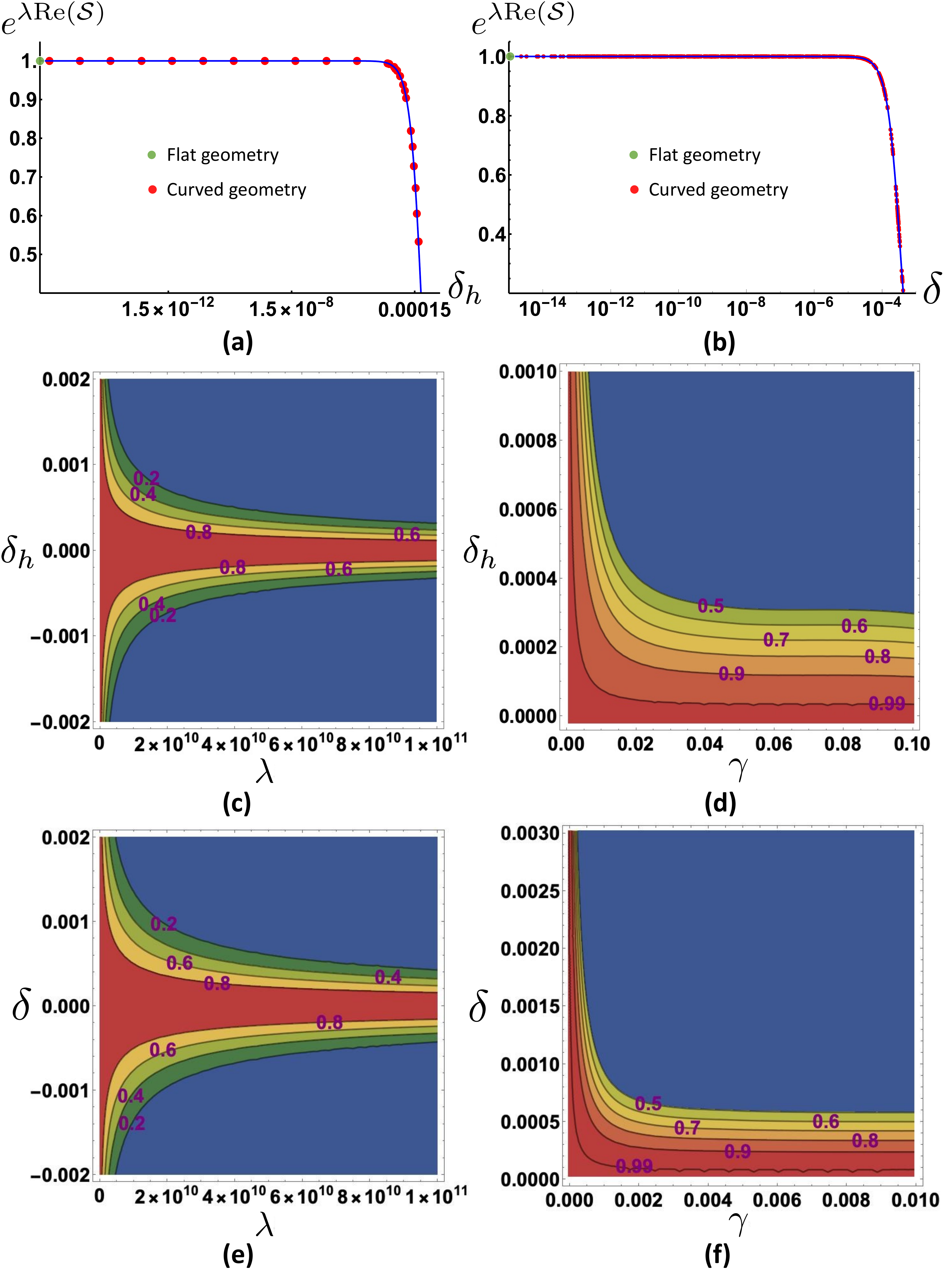}
	\caption{ (a) plots $e^{\lambda\re(\cs)}$ versus the deficit angle $\delta_{h}$ at $\lambda=10^{11}$ and $\gamma=0.1$ in $A(\Delta_3)$, and (b) plots $e^{\lambda\re(\cs)}$ versus the deficit angle $\delta=\sqrt{\frac{1}{10}\sum_{h=1}^{10}\delta_h^2}$ at $\lambda=10^{11}$ and $\gamma=1$ in $\cz_{\sig_{\text 1-5}}$. These 2 plots show the numerical data of curved geometries (red points) and the best fits \eqref{deltaciexp} and \eqref{cspachner} (blue curve). (c) and (d) are the contour plots of $e^{\lambda\re(\cs)}$ as functions of $(\lambda,\delta_h)$ at $\gamma=0.1$ and of $(\g,\delta_{h})$ at $\lambda=5\times10^{10}$ in $A(\Delta_3)$. (e) and (f) are the contour plots of $e^{\lambda\re(\cs)}$ as functions of $(\lambda,\delta)$ at $\gamma=1$ and of $(\g,\delta)$ at $\lambda=5\times10^{10}$ in $\cz_{\sig_{\text{1-5}}}$. They demonstrate the (non-blue) regime of curved geometries where the spinfoam amplitude is not suppressed. 
	}
	\label{thetaS}
\end{figure}


\emph{Asymptotics of $A(\Delta_3)$.}---We firstly focus on a simpler example $A(\Delta_3)$ where $\Delta_3$ contains three 4-simplices and a single internal triangle $h$. All line segments of $\Delta_3$ are at the boundary, and their lengths determine the Regge geometry $\mathbf{g}$ on $\Delta_3$. So $\mathbf{g}$ is fixed by the (Regge-like) boundary data $\{j_b,\xi_{eb}\}$ that uniquely corresponds to the boundary segment-lengths.

Translate the general theory toward \eqref{asymptotics} to $A(\Delta_3)$: $r=\{j_b,\xi_{eb}\}$ is the boundary data. $\mathring{r}=\{\mathring{j}_b,\mathring{\xi}_{eb}\}$ determines the flat geometry $\mathbf{g}(\mathring{r})$ with $\delta_h=0$. $\mathring{x}=\{\mathring{j}_h,\mathring{g}_{ve},\mathring{\bf z}_{vf}\}$ is the real critical point associated to $\mathring{r}$ and endows the orientations $s_v=+1$ to all 4-simplices. $\mathring{r}$, $\mathbf{g}(\mathring{r})$, and $\mathring{x}$ are computed numerically in Appendices \ref{delta3} - \ref{Real critical Point}. The integration domain of $A(\Delta_3)$ is 124 real dimensional. We define local coordinates $x\in\R^{124}$ covering the neighborhood of $\mathring{x}$ inside the integration domain (see Appendix \ref{parametrization}). $S(r,x)$ is the spinfoam action, analytic in the neighborhood of $(\mathring{r},\mathring{x})$. $z\in\C^{124}$ complexifies $x$. $\cs(r,z)$ extends holomorphically $S(r,x)$ to a complex neighborhood of $\mathring{x}$. We only complexify $x$ but do not complexify $r$. We focus on $k_h=0$, since the integrals with $k_h\neq 0$ have no real critical point when $r=\mathring{r}$, and are suppressed even when taking into account complex critical points, as far as $\delta_h$ is not close to $4\pi k_h$.

We vary the length ${l}_{26}$ of the line segment connecting the points 2 and 6, leaving other segment lengths unchanged. A family of (Regge-like) boundary data $r=\mathring{r}+\delta r$ parametrized by ${l}_{26}$ is obtained numerically, and gives the family of curved geometries $\mathbf{g}(r)$ with $\delta_h\neq 0$ (see Appendix \ref{Geometrical perturbations}).

At each $r$, the real critical point is absent. But we find the complex critical point $z=Z(r)$ satisfying $\partial_{z}\cs(r,z)=0$ with high-precision numerics. The details about the numerical solution and error analysis are given in Appendix \ref{Numerical solving complex critical points}. We insert $Z(r)$ into $\cs(r,z)$, and compute numerically the difference between $\cs(r,Z(r))$ and the Regge action $\ci_R$ of the \emph{curved} geometry $\mathbf{g}(r)$:
\be
\delta \ci(r)&=&\cs(r,Z(r))-i\ci_{R}[\mathbf{g}(r)],\label{diff}\\
\text{where}&&\ci_{R}[\mathbf{g}(r)]=\fa_h(r)\delta_h(r)+\sum_b\fa_b(r)\Theta_b(r).
\ee
The areas $\fa_h(r),\fa_b(r)$ and deficit/dihedral angles $\delta_h(r)$,$\Theta_b(r)$ are computed from $\mathbf{g}(r)$. 



We repeat the computation for many $r$ from varying $l_{26}$. The computations give a family of $\delta\ci(r)$. We relate $\delta\ci(r)$ to $\delta_h(r)$ and find the best polynomial fit ({see FIG.\ref{thetaS}(a))
\be
\delta \ci= a_2(\g) \delta_{h}^2+a_3(\g) \delta_{h}^3+a_4 (\g)\delta_{h}^4+ O(\delta_{h}^5),\label{deltaciexp}
\ee
The coefficients $a_i$ at $\g=0.1$ are given in Appendix \ref{NumericalResult}. 

By \eqref{asymptotics}, the dominant contribution from $Z(r)$ to $A(\Delta_3)$ is proportional to $|e^{i\l \cs}|=e^{\l \re(\cs)}\leq 1$. As shown in FIG.\ref{thetaS}(a) and (c), given any finite $\l\gg1$, there are curved geometries with small nonzero $|\delta_h|$ such that $|A(\Delta_3)|$ is the same order of magitude as $|A(\Delta_3)|$ at the flat geometry. The range of $\delta_h$ for non-suppressed $A(\Delta_3)$ is nonvanishing as far as $\l$ is finite. The range of $\delta_h$ is enlarged when $\g$ is small, shown in FIG.\ref{thetaS}(d). 

We remark that the semiclassical behavior of the spinfoam amplitude is given by the $1/\l$ expansion as \eqref{asymptotics} with finite $\l$. It is similar to quantum mechanics where $\hbar$ is finite and the classical mechanics is reproduced by the $\hbar$-expansion. The finite $\l$ leads to the finite range of nonvanishing $\delta_h$.

So far we have considered the real critical point $\mathring{x}$ of the flat geometry with all $s_v=+1$. Given the boundary data $\mathring{r}$, there are exactly 2 real critical points $\mathring{x}$ and $\mathring{x}'$, where $\mathring{x}'$ corresponds to the same flat geometry but with all $s_v=-1$. Other 6 discontinuous orientations (two 4-simplices has plus/minus and the other has minus/plus) do not leads to any real critical point, because they all violates the flatness constraint $\g {\delta}^{s}_h=\g\sum_vs_v\Theta_h(v)=0$. $|{\delta}^{s}_h|$ is not small for the discontinuous orientation, so the contribution to $A(\Delta_3)$ is suppressed even when considering the complex critical point. We focus on the integrals over 2 real neighborhoods $K,K'$ of $\mathring{x},\mathring{x}'$, since the integral outside $K\cup K'$ only gives suppressed contribution to $A(\Delta_3)$ for large $\l$. The above analysis is for the integral over $K$. We carry out a similar analysis for the integral over $K'$. The following asymptotic formula of $A(\Delta_3)$ is obtained with $r=\mathring{r}+\delta r$ of curved geometries $\mathbf{g}(r)$
\be
A(\Delta_3)&=&\lt(\frac{1}{\l}\rt)^{60}\lt[\sn_+ e^{i\l\ci_R[\mathbf{g}(r)]+\l\delta \ci(r)}\rt.\nonumber\\
&&\lt.+\sn_- e^{-i\l\ci_R[\mathbf{g}(r)]+\l \delta \ci'(r)}\rt]\lt[1+O(1/\l)\rt].\label{asymp}
\ee
up to an overall phase. 2 complex critical points in complex neighborhoods of $\mathring{x},\mathring{x}'$ contribute dominantly and give respectively 2 terms, with phase plus or minus the Regge action of the curved geometry $\mathbf{g}(r)$ plus curvature corrections $\delta\ci(r)$ in \eqref{deltaciexp} and $\delta\ci'(r)=\delta\ci(r)^*|_{\delta_h\to -\delta_h}$. $\sn_\pm$ are proportional to $[\det(-\delta^2_{z,z}\cs/2\pi)]^{-1/2}$ evaluated at these 2 complex critical points.

As an example of the suspected \emph{cosine problem} \cite{Engle:2011pmf}, there has been the guess $A(\Delta_3)\sim (\sn_1 e^{i\l \ci_R}+\sn_2 e^{-i\l \ci_R})^3$ (each factor is from the vertex amplitude, see e.g. \cite{Dona:2020tvv}) whose expansion gives 8 terms corresponding to all possible orientations. But Eq.\eqref{asymp} demonstrates that $A(\Delta_3)$ only contain 2 terms corresponding to the continuous orientations. The cosine problem is relaxed.

\emph{Internal segments.}---Let us consider a triangulation $\ck$ that have $M>0$ internal line segments, in contrast to $\Delta_3$ where all segments are at the boundary. The internal segment is labelled by $I$. We consider a real critical point $\{\mathring{j}_{h},\mathring{g}_{ve},\mathring{\bf z}_{vf}\}$ of flat geometry on $\ck$. To generalize the above analysis, we make a change of variable in \eqref{integralFormAmp} by replacing $M$ internal areas $j_{h_o}$ ($h_o=1,\cdots,M$) to all internal segment-lengths $l_I$ and denoting other $j_h$ by $j_{\bar{h}}$ ($\bar{h}=1,\cdots,F-M$ where $F$ is the number of internal triangles). This change of variable is done locally by inverting Heron's formula in a neighborhood of $\{\mathring{j}_{h_o}\}$ of $\{\mathring{j}_h,\mathring{g}_{ve},\mathring{\bf z}_{vf}\}$, and $\rmd^F j_h=\cj_l\rmd^M l_I\,\rmd^{F-M} j_{\bar{h}}$ where $\cj_l$ is the jacobian. Therefore
\be
A(\ck)&=&\int\prod_{I=1}^M \rmd l_I \cz_{\ck}\lt(l_I\rt),\label{integralFormAmp2}\\
\cz_{\ck}\lt(l_I\rt)&=&\sum_{\{k_h\}}\int \prod_{\bar{h}}\mathrm{d} j_{\bar h}\, \prod_h\lt(2\l \bm{d}_{\l j_{h}}\rt)\int [\rmd g\rmd \mathbf{z}] e^{\lambda S^{(k)}}\cj_l,\nonumber
\ee
 We apply the procedure of \eqref{asymptotics} to $\cz_{\ck}$. The integrals in $\cz_{\ck}$ have external parameters $r\equiv\{l_I,j_b,\xi_{eb}\}$ including not only the boundary data but also internal segment-lengths. $\mathring{r}\equiv\{\mathring{l}_I,\mathring{j}_b,\mathring{\xi}_{eb}\}$ gives internal and boundary segment-lengths of the flat geometry $\mathbf{g}(\mathring{r})$ on $\ck$ ($\mathring{j}_b,\mathring{\xi}_{eb}$ determine boundary segment-lengths). Focusing on $k_h=0$, $\mathring{x}=\{\mathring{j}_{\bar h},\mathring{g}_{ve},\mathring{\bf z}_{vf}\}$ is the real critical point of $\mathbf{g}(\mathring{r})$ in the integral. There are local coordinates $x\in\R^N$ covering a neighborhood $K$ of $\mathring{x}$. We express the spinfoam action as $S(r,x)$ and analytic continue to $\cs(r,z)$ where $z\in\C^N$. We fix the boundary data and deform $l_I=\mathring{l}_I+\delta l_I$ so that $r=\{{l}_I,\mathring{j}_b,\mathring{\xi}_{eb}\}\equiv r_l$ give flat and curved geometries $\mathbf{g}(r_l)$. As in \eqref{asymptotics}, the dominant contribution to $\cz_{\ck}(l_I)$ from the complex critical point $Z(r_l)$ reads
\be
\left(\frac{1}{\lambda}\right)^{\frac{N}{2}-2F}\!\!\!\int\prod_{I=1}^M \rmd l_I \sn_l\,{e^{\lambda \mathcal{S}(r_l,Z(r_l))}} \lt[1+O(1/\l)\rt],\label{pathintegral}
\ee
which reduces $A(\ck)$ to the integral over geometries $\mathbf{g}(r_l)$ in the neighborhood of $(\mathring{r},\mathring{x})$. $\sn_l$ is proportional to $\prod_h\lt(4 j_{h}\rt)\cj_l[\det(-\delta^2_{z,z}\cs/2\pi)]^{-1/2}$ at $Z(r_l)$. $\mathcal{S}(r_l,Z(r_l))$ is the Regge action of $\mathbf{g}(r_l)$ plus the curvature correction of $O(\delta_h^2)$, similar to \eqref{asymp}. This is confirmed numerically by the following example.

\emph{1-5 Pachner move.}--- $\sig_{\text{1-5}}$ is the complex of the 1-5 Pachner move refining a 4-simplex into five 4-simplices. $\sig_{\text{1-5}}$ has $F=10$ internal triangles $h$ and $M=5$ internal segments $I=1,\cdots,5$ (see FIG.\ref{pic}(b)). The integral in $\cz_{\sig_{\text{1-5}}}(l_I)$ is 195 real dimensional. Fixing the boundary data, $r_l$ and curved geometries $\mathbf{g}(r_l)$ on $\sig_{\text{1-5}}$ are parametrized by five $l_I$. We randomly sample $l_I$ of flat and curved geometries and compute numerically the real and complex critical points $Z(r_l)$. The computation is similar to $A(\Delta_3)$. The numerical result of $|e^{i\l \cs}|=e^{\l\mathrm{Re}(\cs)}$ is presented in FIG.\ref{thetaS} (b), (e), and (f), which demonstrate curved geometries with small $|\delta_h|$ do not lead to the suppression of $\cz_{\sig_{\text{1-5}}}(l_I)$. Moreover $\cs(r_l,Z(r_l))$ in \eqref{pathintegral} is numerically fit by:
\be
\!\!\!\!\!\cs(r_l,Z(r_l))=-i\ci_R[\mathbf{g}(r_l)]-{a_2(\g)}\delta(r_l)^2+O(\delta^3),\label{cspachner}
\ee
where $\delta(r_l)=\sqrt{\frac{1}{10}\sum_{h=1}^{10}\delta_h(r_l)^2}$ and $a_2=-0.033i +8.88\times 10^{-5}$ at $\g=1$. $\ci_R[\mathbf{g}(r_l)]$ is the Regge action of $\mathbf{g}(r_l)$. Some more details are given in Appendix \ref{Pachner moves in a 4d Lorentzian Spinfoam model}.


\emph{Discussion.}--- Our results resolve the flatness problem by demonstrating explicitly the curved Regge geometries emergent from the large-$j$ EPRL spinfoam amplitudes. The curved geometries correspond to complex critical points that are away from the real integration domain. They give non-suppressed $e^{\l \re(\cs)}$ and satisfy the bound ${\mathrm{Re}(a_2(\g))} \delta^2\lesssim1/\l$, if we consider the examples \eqref{deltaciexp} and \eqref{cspachner} neglecting $O(\delta^3)$. This bound is consistent with earlier results in e.g. \cite{Han:2013hna,Asante:2020qpa}, although this bound should be corrected when taking into account $O(\delta_h^3)$ in \eqref{deltaciexp} and \eqref{cspachner}. The similar bound should be valid to the spinfoam amplitude in general.

All resulting curved geometries are of small deficit angles $\delta_h$. The large-$j$ spinfoam amplitude is still suppressed for geometries with larger $\delta_h$ violating the above bound. This is not a problem for the semiclassical analysis. Indeed, non-singular classical spacetime geometries are smooth with vanishing $\delta_h$. In order to well-approximating smooth geometries by Regge geometries, the triangulation must be sufficiently refined, and all $\delta_h$'s must be small.

The 1-5 pachner move is an elementary step in the triangulation refinement. Our results provide a new routine for analyzing the triangulation refinement in the spinfoam model. This should closely relate to the spinfoam renormalization \cite{Bahr:2016hwc,Delcamp:2016dqo} with the goal to resolve the issue of triangulation-dependence of the spinfoam theory.

\begin{acknowledgements}


The authors acknowledge Jonathan Engle and Wojciech Kaminski for helpful discussions. M.H. receives support from the National Science Foundation through grant PHY-1912278. Z.H. is supported by Xi De post-doc funding from State Key Laboratory of Surface Physics at Fudan University.

\end{acknowledgements}


\onecolumngrid




\section{Spinfoam amplitude and Poisson summation}\label{Spinfoam amplitude, Poisson resummation formula, and parametrization}

The Lorentzian EPRL spinfoam amplitude on the simplicial complex $\ck$ has the following integral expression:
\be
A(\ck)&=&\sum_{\{j_{h}\}}^{j^{\rm max}}\prod_h \bm{d}_{j_h}\int [\rmd g\rmd\mathbf{z}]\, e^{S
\left(j_{h}, g_{v e}, \mathbf{z}_{vf};j_b,\xi_{eb}\right)},\quad [\rmd g\rmd \mathbf{z}]=\prod_{(v, e)} \mathrm{d} g_{v e} \prod_{(v,f)} \mathrm{d}\O_{\mathbf{z}_{v f}}, \label{amplitude1}
\ee
where
\be
S
&=&\sum_{e'}j_hF_{(e',h)}+\sum_{(e,b)}j_bF^{in/out}_{(e,b)}+\sum_{(e',b)}j_bF^{in/out}_{(e',b)},\label{SjFjF} 
\\
F_{(e,b)}^{out}&=&2 \ln\dfrac{\left\langle Z_{v e b},\xi_{e b}\right\rangle}{\left\| Z_{v e b}\right\|}+i\g \ln \left\| Z_{v e b}\right\|^2,\\
F_{(e,b)}^{in}&=& 2 \ln \dfrac{\left\langle\xi_{e b}, Z_{v' e b}\right\rangle}{\left\| Z_{v' e b}\right\|}-i\g \ln \left\| Z_{v' e b}\right\|^2,\\
F_{(e',f)}&=&2 \ln \dfrac{\left\langle Z_{v e' f}, Z_{v^{\prime} e' f}\right\rangle}{\left\| Z_{v e' f}\right\|\left\| Z_{v^{\prime} e' f}\right\|} + i\g \ln\frac{\left\| Z_{v e' f}\right\|^2}{\left\| Z_{v^{\prime} e' f}\right\|^2}.
\ee
$Z_{vef}=g^\dagger_{ve}\mathbf{z}_{vf}$ and $f=h$ or $b$. $e$ and $e'$ are boundary and internal tetrahedra respectively. Introducing the dual complex $\ck^*$, the orientation of the face $f^*$ dual to $f$ induces $\partial f^*$'s orientation that is outgoing from the vertex dual to $v$ and incoming to another vertex dual to $v'$. The logarithms are fixed to be the principle value.

We have assume the sum over internal $j_h\in\mathbb{N}_0/2$ is bounded by $j^{\rm max}$. For some internal triangles $h$, $j^{\rm max}$ is determined by boundary spins $j_b$ via the triangle inequality, or $j^{\rm max}$ is an IR cut-off in case of the bubble divergence.

We would like to change the sum over $j_h$ to the integral, preparing for the stationary phase analysis. The idea is to apply the Poisson summation formula. Firstly, we replace each $\bm{d}_{j_h}$ by a smooth compact support function $\t_{[-\epsilon,j^{\rm max}+\epsilon]}(j_h)$ satisfying
\be
\t_{[-\epsilon,j^{\rm max}+\epsilon]}( j_h)=\bm{d}_{j_h},\quad j_h\in[0,j^{\rm max}]\quad \text{and}\quad \t_{[-\epsilon,j^{\rm max}+\epsilon]}(j_h)=0,\quad j_h\not\in[-\epsilon,j^{\rm max}+\epsilon],
\ee
for any $0<\epsilon<1/2$. This replacement does not change the the value of the amplitude $A(\ck)$, but makes the summand of $\sum_{j_h}$ smooth and compact support in $j_h$. Applying Poisson summation formula
\[
\sum_{n\in\Z} f(n)=\sum_{k \in \mathbb{Z}} \int_\R \mathrm{d} n f(n) \,{e}^{2\pi i k n},
\]
the discrete sum over $j_h$ in $A(\ck)$ becomes the integral. Therefore,
\be
A(\ck)&=&\sum_{\{k_h\in\mathbb{Z}\}} \int_\R
\prod_h\mathrm{d} j_{h}\prod_h 2\t_{[-\epsilon,j^{\rm max}+\epsilon]}(j_h)\int [\rmd g\rmd \mathbf{z}]\, e^{S^{(k)}},\quad S^{(k)}=S+4\pi i \sum_h j_h k_h.\label{integralFormAmp1}
\ee
To probe the large-$j$ regime, we scale boundary spins $j_b\to \l j_b$ with any $\l\gg1$, and make the change of variables $j_h\to\l j_h$. We also scale $j^{\rm max}$ by $j^{\rm max}\to \l j^{\rm max}$. Then, $A(\ck)$ is given by  
\be
A(\ck)&=&\sum_{\{k_h\in\mathbb{Z}\}}\int_\R
\prod_h\mathrm{d} j_{h}\prod_h 2 \l\,\t_{[-\epsilon,\l j^{\rm max}+\epsilon]}(\l j_h)\int [\rmd g\rmd \mathbf{z}]\, e^{\l S^{(k)}},\quad S^{(k)}=S+4\pi i \sum_h j_h k_h,\label{integralFormAmp22}
\ee
which is used in our discussion.

\section{The spinfoam amplitude $A(\Delta_3)$}

\subsection{The flat geometry on $\Delta_3$} \label{delta3}
The $\Delta_3$ triangulation is made by three 4-simplices sharing a common triangle (see FIG.\ref{pic}(a)). $\Delta_3$ has 18 boundary triangles and one internal triangle (the red triangle in FIG.\ref{pic}(a)). All line segments of $\Delta_3$ are at the boundary, and the segment-lengths $l_{ab}(a\neq b = 1,2,3,4,5,6)$ determine the Regge geometry $\textbf{g}(r)$ ($\textbf{g}(r)$ does not contain the information of the 4-simplex orientations)

The dual cable diagram for the $\Delta_3$ triangulation is represented in FIG.\ref{4complex}(a). Each box in FIG.\ref{4complex} carries group variables $g_{a}\in\Slc$ \interfootnotelinepenalty=10000\footnote{For convenience, the indexes of group variables in FIG. \ref{4complex}(a) are $a=1,2,3...,15$, the corresponding tetrahedra $e$ are labeled by the number circles in FIG.\ref{pic}(a). The correspondence are: $g_1\rightarrow e_{2,3,4,5}$, $g_2\rightarrow e_{1,2,4,5}$, $g_3\rightarrow e_{1,2,3,4}$, $g_4\rightarrow e_{1,3,4,5}$, $g_5\rightarrow e_{1,2,3,5}$, $g_6\rightarrow e_{1,2,3,5}$, $g_7\rightarrow e_{1,2,5,6}$, $g_8\rightarrow e_{1,3,5,6}$, $g_9\rightarrow e_{1,2,3,6}$, $g_{10}\rightarrow e_{2,3,5,6}$, $g_{11}\rightarrow e_{1,3,5,6}$, $g_{12}\rightarrow e_{1,3,4,5}$, $g_{13}\rightarrow e_{1,4,5,6}$, $g_{14}\rightarrow e_{1,3,4,6}$, $g_{15}\rightarrow e_{3,4,5,6}$.}. Each strand carries an $\Su$ spin $j_{a,b}$ where $a,b$ corresponds to 2 different tetrahedra sharing the same 4-simplex. We have the identification $j_{a,b}$'s along the same strand, e.g. $j_{2,5}=j_{6,7}$ along the pink strand. The red strands is dual to the common triangle shared by three 4-simplices. We use $j_h$ to denote the spin $j_{4,5}=j_{6,8}=j_{11,12}$ of the internal triangle. The circles at the ends of the strands represent the $\Su$ coherent states. 

\begin{figure}[h]	
	\centering
	\includegraphics[scale=0.37]{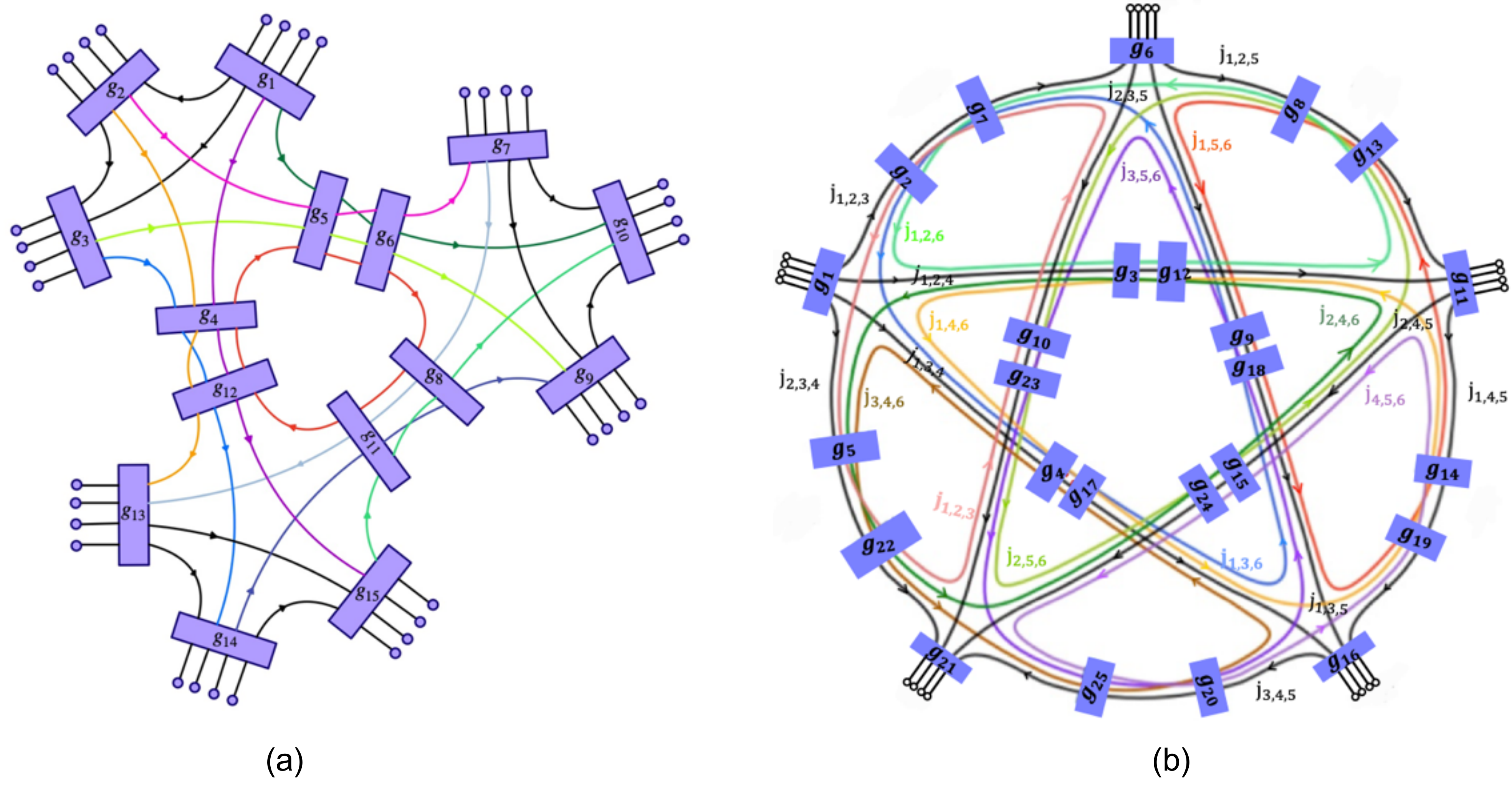}
	\caption{(a). The dual cable diagram of the $\Delta_3$ spinfoam amplitude: The boxes correspond to tetrahedra carrying $g_{a} \in \Slc$. The strands stand for triangles carrying spins $j_{f}$. The strand with same color belonging to different dual vertex corresponds to the triangle shared by the different 4-simplices. The circles as the endpoints of the strands carry boundary states $|j_b,\xi_{eb}\rangle$. The arrows represent orientations. This figure is adapted from \cite{Dona:2020yao}. (b). The dual cable diagram of the 1-5 pachner move amplitude. The internal faces are colored loops carrying internal spins $j_h$. The boundary faces are black strands carrying boundary spins $j_b$. The arrows represent orientations. This figure is adapted from \cite{Banburski:2014cwa}. }
	\label{4complex}
\end{figure} 

We firstly construct the flat Regge geometry on $\Delta_3$, in order to obtain the corresponding boundary data $\mathring{r}=\lbrace \mathring{j}_b,\mathring{\xi}_{eb}\rbrace$ and compute the associated real critical point $\mathring{x}$. We set the 6 points of $\Delta_3$ in $\R^4$ as 
{\small
\be
& P_{1}=(0,0,0,0),P_{2}=\left(0,-2 \sqrt{10} / 3^{3 / 4},-\sqrt{5} / 3^{3 / 4},-\sqrt{5} / 3^{1 / 4}\right),\, P_{3}=\left(0,0,0,-2 \sqrt{5} / 3^{1 / 4}\right),\\ &P_{4}=\left(-3^{-1 / 4} 10^{-1 / 2},-\sqrt{5 / 2} / 3^{3 / 4},-\sqrt{5} / 3^{3 / 4},-\sqrt{5} / 3^{1 / 4}\right), P_{5}=\left(0,0,-3^{1 / 4} \sqrt{5},-3^{1 / 4} \sqrt{5}\right), \,P_6=\left(0.90, 2.74, -0.98, -1.70 \right)\nonumber.\label{P6}
\ee}
The 4-simplex with points $(12345)$ has the same 4-simplex geometry as in \cite{Han:2020fil,Dona:2019dkf}. We choose $P_6$ in (\ref{P6}) so that we have the length symmetry $l_{12}=l_{13}=l_{15}=l_{23}=l_{25}=l_{35}\approx 3.40$, $l_{14}=l_{24}=l_{34}=l_{45}\approx 2.07$, $l_{16}=l_{36}=l_{56}\approx 3.25$, $l_{26}\approx5.44$ and $l_{46}\approx 3.24$.

All tetrahedra and triangles are space-like. The tetrahedron 4-d normal vectors $N_a (a=1,2,...,15)$ are determined by the triple product of three segment-vectors $l^\mu_{1},l^\mu_2,l^\mu_3$ (the segment-vectors are given by $P_i^\mu-P_j^\mu$) along three line-segments labeled by $1,2,3$ adjacent to a common point
\begin{equation}
(N_{a})_\mu=\frac{\epsilon_{\mu\nu\rho\sig} l_{a 1}^{\nu} l_{a 2}^{\rho} l_{a 3}^{\sig}}{\left\|\epsilon_{\mu\nu\rho\sig} l_{a 1}^{\nu} l_{a 2}^{\rho} l_{a 3}^{\sig}\right\|}, \label{4dnormals}
\end{equation}
where the norms $\| \cdot \|$ is given by the Minkowski metric $\eta=\text{diag}(-,+,+,+)$, and $\epsilon_{\mu\nu\rho\sig}$ follows the convention $\epsilon_{0123}=1$. We list below the 4-d normals $(N_{a})_\mu$ of the tetrahedra in each 4-simplex: 
\begin{itemize}
	\item The first 4-simplex with points 12345:
	\be\small
	&N_1=\left(1.07, -0.12, -0.17, -0.30\right),\,N_2=\left(1.07, -0.12, -0.17, 0.30\right),\nonumber\\
	&N_3=\left(1.07, -0.12, 0.35, 0\right),\, N_4=\left(1.07, 0.37, 0, 0\right),\,N_5=\left(-1, 0, 0, 0\right).
	\ee
	\item The second 4-simplex  with points 12456:
	\be
	&N_6=\left(1,0,0,0\right), \,N_7=\left(-1.15, -0.19, -0.26, 0.46\right),\,N_8=\left(1.06, 0.35, 0,0\right),\nonumber\\ &N_9=\left(-1.15, -0.19, 0.53, 0\right),\, N_{10}=\left(-1.15, -0.19, -0.26, -0.46\right).
	\ee
	\item The third 4-simplex  with points 13456:
	\be
	&N_{11}=\left(-1, 0.02, 0, 0\right),\, N_{12}=\left(-1,0,0,0\right),\,N_{13}=\left(1, -0.02, -0.01, 0.01\right),\nonumber\\ &N_{14}=\left(1, -0.02, 0.01, 0\right),\,N_{15}=\left(1, -0.02, -0.01, -0.01\right).
	\ee
\end{itemize}
The triangles of within a 4-simplex are classified into two categories \cite{Barrett:2009mw}: The triangle corresponds to the \emph{thin wedge} if the inner product of normals is positive; 
The triangle corresponds to \emph{thick wedge} if the inner product of normals is negative. The dihedral angle $\theta_{a,b}$ are determined by:
\be
&\text{thin wedge:}\qquad N_a\cdot N_b&=\cosh\theta_{a,b},\nonumber\\
&\text{thick wedge:}\qquad N_a\cdot N_b&=-\cosh\theta_{a,b}.\label{dihedral}
\ee
where the inner product is defined by $\eta$. Then we check the deficit angle $\delta_h$ associated to the shared triangle $h$ 
\be
0=\delta_h=\theta_{4,5}+\theta_{6,8}+\theta_{11,12}\approx 0.36-0.34-0.02,\label{deficit0}
\ee
which implies the Regge geometry is flat. \\

To determine the 3-d normals of triangles, we proceed with a similar method as in \cite{Dona:2019dkf}. To transform all 4-d normals to $t^\mu=(1,0,0,0)$, we use the following pure boost $\Lambda_a\in O(1,3)$:
\be
	(\Lambda_{a})^{\nu}{}_ {\rho}=\sigma\eta^{\nu}_{\rho}+\frac{\sigma}{1-\sigma N_a\cdot t}\bigg(N^{\nu}_a N_{a\rho}+t^{\nu}t_{\rho}+\sigma N^{\nu}_a t_{\rho}-(1-2\sigma N_a\cdot t) \sigma t^{\nu} N_{a\rho} \bigg), \label{O13}
\ee
where $\sigma=1$ for $N_{a0}>0$ or $\sigma=-1$ for $N_{a0}<0$. Then, the 3-d normals $\vec{\fn}_{a,b}$ can be expressed by $\Lambda_a$ and 4-d normals:
\begin{equation}
\fn_{a,b}:=(0,\vec{\fn}_{a,b})=(\Lambda_{a})^{\nu}{}_ {\rho} \frac{N^{\rho}_b+N^{\rho}_a\left(N_b\cdot N_a\right)}{\sqrt{(N_b\cdot N_a)^2-1}}.\label{nab}
\end{equation}
Here, $\vec{\fn}_{a,b}$ are the outward normals of the triangles in the tetrahedron $a$, then the inward normals are $-\vec{\fn}_{a,b}$. We associate $\vec{n}_{a,b}=\vec{\fn}_{a,b}$ (or $-\vec{\fn}_{a,b}$) to a strand oriented outward from (or inward to) the box labelled by $g_a$. The data of $\vec{\fn}_{a,b}$ can be found in the Mathematica notebook \cite{qudx_Delta3.org}.

The spinors $\xi_{eb}$ in Eq.(\ref{SjFjF}) relate to $\vec{n}_{a,b}$ by $\vec{n}_{a,b}=\langle \xi_{a,b},\vec{\sigma} \xi_{a,b}\rangle$. We use the following rule to convert a unit 3-vector to a normalized spinor (by fixing the phase convention):
\be
\vec{n}_{a,b}=\left(x,y,z\right) \quad \rightarrow  \quad \xi_{a,b}=\frac{1}{\sqrt{2}}\left(\sqrt{1+z},\,\frac{x+iy}{\sqrt{1+z}}\right). \label{xi}
\ee 
The data for $j_{a,b}, \xi_{a,b}$ are listed in Tables \ref{bdata1}, \ref{bdata2}, \ref{bdata3}. In these tables, $j_{a,b}, \xi_{a,b}$ for the internal face are labeled in the bold text, and the others are the boundary data. We denote the boundary data in these tables by $\mathring{r}=(\mathring{j}_b,\mathring{\xi}_{eb})$. 

\begin{table}[H] 
	\centering \caption{Geometry data $\mathring{j}_{a,b},\mathring{\xi}_{a,b}$ for 1st 4-simplex with points 12345}\label{bdata1}
	\scalebox{0.94}{
	\begin{tabular}{|c|c|c|c|c|c|} 
		\hline
		\diagbox{\small{a}}{$\mathring{\xi}_{a,b}$}{\small{b}}&1&2&3&4&5\\
		\hline
		1&\diagbox{}{}&(1.,0.01 + 0.01i)&(0.87,0.01+0.49i)&(0.87,0.46+0.17i)&(0.3, -0.55-0.78i)\\
		\hline
		2&(1,-0.01,-0.01i)&\diagbox{}{}&(0.49,0.02+0.87i)&(0.49,0.82+0.31i)&(0.95,-0.17-0.25i)\\
		\hline
		3&(0.86,-0.01+0.51i)&(0.51,-0.02+0.86i)&\diagbox{}{}&(0.71,0.56-0.43i)&(0.71,-0.24+0.67i)\\
		\hline
		4&(0.86,0.48+0.16i)&(0.51,0.82+0.27i)&(0.71,0.59-0.39i)&\diagbox{}{}&$\boldsymbol{(0.71,0.71)}$\\
		\hline
		5&(0.3,-0.55-0.78i)&(0.95,-0.17-0.25i)&(0.71,-0.24+0.67i)&$\boldsymbol{(0.71,0.71)}$&\diagbox{}{}\\
		\hline
	\end{tabular}	
	
	\begin{tabular}{|c|c|c|c|c|c|}
		\hline
		\diagbox{\small{a}}{$\mathring{j}_{a,b}$}{\small{b}}&1&2&3&4&5\\
		\hline
		1&\diagbox{}{}&2&2&2&5\\
		\hline
		2&\diagbox{}{}&\diagbox{}{}&2&2&5\\
		\hline
		3&\diagbox{}{}&\diagbox{}{}&\diagbox{}{}&2&5\\
		\hline
		4&\diagbox{}{}&\diagbox{}{}&\diagbox{}{}&\diagbox{}{}&\textbf{5}\\
		\hline
	\end{tabular}	
	}
\end{table}

\begin{table}[H]
	\centering \caption{Geometry data $\mathring{j}_{a,b},\mathring{\xi}_{a,b}$ for 2nd 4-simplex with points 12456}\label{bdata2}
	\scalebox{0.89}
	{
	\begin{tabular}{|c|c|c|c|c|c|} 
		\hline
		\diagbox{\small{a}}{$\mathring{\xi}_{a,b}$}{\small{b}}&6&7&8&9&10\\
		\hline
		6&\diagbox{}{}&(0.95,-0.17-0.25i)&$\boldsymbol{(0.71,0.71)}$&(0.71,-0.24+0.67i)&(0.3,-0.55-0.78i)\\
		\hline
		7&(0.95,-0.17-0.25i)&\diagbox{}{}&(0.29,-0.47+0.83i)&(0.88,-0.02-0.48i)&(1,-0.02-0.03i)\\
		\hline
		8&$\boldsymbol{(0.71,0.71)}$&(0.31,-0.57+0.76i)&\diagbox{}{}&(0.71,0.25+0.66i)&(0.31,0.57-0.76i)\\
		\hline
		9&(0.71,-0.24+0.67i)&(0.85,0.02-0.52i)&(0.71,0.19+0.68i)&\diagbox{}{}&(0.85,-0.02+0.52i)\\
		\hline
		10&(0.3,-0.55-0.78i)&(1,0.02+0.03i)&(0.29,0.47-0.83i)&(0.88,0.02+0.48i)&\diagbox{}{}\\
		\hline
	\end{tabular}
	
	\begin{tabular}{|c|c|c|c|c|c|}
		\hline
		\diagbox{\small{a}}{$j_{a,b}$}{\small{b}}&6&7&8&9&10\\
		\hline
		6&\diagbox{}{}&5&\textbf{5}&5&5\\
		\hline
		7&\diagbox{}{}&\diagbox{}{}&4.71&5.19&5.19\\
		\hline
		8&\diagbox{}{}&\diagbox{}{}&\diagbox{}{}&4.71&4.71\\
		\hline
		9&\diagbox{}{}&\diagbox{}{}&\diagbox{}{}&\diagbox{}{}&5.19\\
		\hline
	\end{tabular}
	}
\end{table}

\begin{table}[H]
	\centering\caption{Geometry data $\mathring{j}_{a,b},\mathring{\xi}_{a,b}$ for 3rd 4-simplex with points 13456}\label{bdata3}
	\scalebox{0.89}
	{
	\begin{tabular}{|c|c|c|c|c|c|}
		\hline
		\diagbox{\small{a}}{$\mathring{\xi}_{a,b}$}{\small{b}}&11&12&13&14&15\\
		\hline
		11&\diagbox{}{}&$\boldsymbol{(0.71,0.71)}$&(0.31,-0.57+0.76i)&(0.71,0.25+0.66i)&(0.31,0.57-0.76i)\\
		\hline
		12&$\boldsymbol{(0.71,0.71)}$&\diagbox{}{}&(0.51,0.82+0.27i)&(0.71,0.59-0.39i)&(0.86,0.48+0.16i)\\
		\hline
		13&(0.31,-0.57+0.76i)&(0.51,0.82+0.27i)&\diagbox{}{}&(0.5,0.87i)&(0,0.95+0.31i)\\
		\hline
		14&(0.71,0.25+0.66i)&(0.71,0.59-0.39i)&(0.5,0.87i)&\diagbox{}{}&(0.5,-0.87i)\\
		\hline
		15&(0.31,0.57-0.76i)&(0.86,0.48+0.16i)&(0,-0.95-0.31i)&(0.5,-0.87i)&\diagbox{}{}\\
		\hline
	\end{tabular}
	
	\begin{tabular}{|c|c|c|c|c|c|}
		\hline
		\diagbox{\small{a}}{$\mathring{j}_{a,b}$}{\small{b}}&11&12&13&14&15\\
		\hline
		11&\diagbox{}{}&\textbf{5}&4.71&\diagbox{}{}&\diagbox{}{}\\
		\hline
		12&\diagbox{}{}&\diagbox{}{}&2&2&2\\
		\hline
		13&\diagbox{}{}&\diagbox{}{}&\diagbox{}{}&3.18&3.18\\
		\hline
		14&4.71&\diagbox{}{}&\diagbox{}{}&\diagbox{}{}&3.18\\
		\hline
		15&4.71&\diagbox{}{}&\diagbox{}{}&\diagbox{}{}&\diagbox{}{}\\
		\hline
	\end{tabular}
	}
\end{table}
Once the flat geometry data $\mathring{\xi}_{a,b}$ and $\mathring{j}_{a,b}$ are constructed, we are ready to obtain the real critical points $\mathring{x}=(\mathring{j}_h,\mathring{g}_{a},\mathring{\textbf{z}}_{a,b})$ by solving the critical point equations (\ref{eom1}) and \eqref{eom2}. Here $\mathring{j}_h=\mathring{j}_{4,5}=\mathring{j}_{6,8}=\mathring{j}_{11,12}=5$ is the same as the area of $h$.

\subsection{The real critical point}\label{Real critical Point}

The solution of the critical point equations Eq.(\ref{eom1}) relates to the Lorentzian Regge geometry, as described in \cite{Han:2011re,Barrett:2009mw}. $\mathring{g}_{a}$ relates to the Lorentzian transformation acting on each tetrahedron and gluing them together to form the $\Delta_3$ triangulation. The general form of $\mathring{g}_a$ can be expressed by:
\be
\mathring{g}_a=\exp\left(\theta_{\text{ref},\,a}\vec{n}_{\text{ref},\,a}\cdot\frac{\vec{\sigma}}{2} \right), \label{group}
\ee
where $\theta_{\text{ref},\,a}$ is the dihedral angle which is defined in Eq. (\ref{dihedral}), $\vec{\sigma}$ are the Pauli matrices, and $\text{ref}=5,6,12$ are the reference tetrahedra, whose 4-d normals equal $\pm t$. The data for 3-d normals $\vec{n}_{\text{ref},\,a}$ can be found in Mathematica notebook \cite{qudx_Delta3.org}. The Spinfoam action has the following continuous gauge freedom:
\bi
\item At each $v$, there is the $\Slc$ gauge freedom  $g_{ve}\mapsto x_v^{-1}g_{ve}$, $ \textbf{z}_{vf}\mapsto x_v^{\dagger}\textbf{z}_{vf}$, $x_v\in\Slc$. We fix $g_{a}$ to be a constant $\Slc$ matrix for $a=1,10,15$ \footnote{The choice of $a=1,10,15$ for the $\Slc$ gauge fixing is different from the $\mathrm{ref}=5,6,12$, because we would like to apply the $\Slc$ and $\Su$ gauge fixings to different sets of $g_a$'s.}. The amplitude is independence of the choices of constant matrices.
\item At each $e$, there is the $\Su$ gauge freedom: $g_{v'e}\mapsto g_{v'e}h_e^{-1}$, $g_{ve}\mapsto g_{ve}h_e^{-1}$, $h_e\in\Su$. To remove the gauge freedom, we set one of the group element $g_{v'e}$ along the edge $e$ to be the upper triangular matrix. Indeed, any $g\in\Slc$ can be decomposed as $g=kh$ with $h\in\Su$ and $k\in K$, where $K$ is the subgroup of upper triangular matrices:
\begin{equation}
K=\left\lbrace k=\left(\begin{matrix}
\lambda^{-1}&\mu\\0&\lambda
\end{matrix}\right),\ \lambda\in\R\setminus\{0\},\ \mu\in\mathbb{C}\right\rbrace.
\end{equation}  
We use the gauge freedom to set $g_{v'e}\in K$.
On $\Delta_3$, we fix the group element $g_a$ for the bulk tetrahedra $a=5, 8, 12$ to be the upper triangular matrix. 

\item $\textbf{z}_{vf}$ can be computed by $g_{ve}$ and $\xi_{ef}$ up to a complex scaling: $\textbf{z}_{vf}\propto_\mathbb{C}\left(g_{ve}^\dagger\right)^{-1}\xi_{ef}$. Each $\textbf{z}_{vf}$ has the scaling gauge freedom $\textbf{z}_{vf}\mapsto\lambda_{vf} \textbf{z}_{vf}$, $\lambda_{vf}\in\mathbb{C}$.  We fix the gauge by setting the first component of $\textbf{z}_{vf}$ to 1. Then, the real critical point $\mathring{\textbf{z}}_{vf}$ is in the form of $\mathring{\textbf{z}}_{vf}=\left(1, \mathring{\a}_{vf}\right)^T$, where $\mathring{\a}_{vf}\in\C$.
\ei

By Eq.(\ref{group}) and the gauge fixing for $g_{ve},\, \textbf{z}_{vf}$, we obtain the numerical results of the real critical points $(\mathring{j}_h,\mathring{g}_a,\,\mathring{\textbf{z}}_{a,b})$ corresponding to the flat geometry and all $s_v=+1$. $j_h=5$ as the area of the internal triangle. The numerical data of $\mathring{g}_a,\,\mathring{\textbf{z}}_{a,b}$ are shown in Table \ref{tab:ga1}, \ref{tab:ga2} and \ref{tab:ga3}.
\begin{table}[H]
	\centering\caption{The real critical point $\mathring{g}_{a},\, \mathring{\textbf{z}}_{a,b}$ for the 1st 4-simplex with points 12345.}\label{tab:ga1}
	\scalebox{0.64}
	{
	\begin{tabular}{|c|c|c|c|}
		\hline
		\small{a}&1&2&3\\
		\hline
		$\mathring{g}_{a}$ &$\left(\begin{matrix}
		0.87&-0.06+0.09i\\	-0.06-0.09i&1.16
		\end{matrix}\right)$&$\left(\begin{matrix}
		1.16 &-0.06+0.09i\\	-0.06-0.09i&0.87
		\end{matrix}\right)$&$\left(\begin{matrix}
		1.02&-0.06-0.17i\\	-0.06+0.17i&1.02
		\end{matrix}\right)$\\
		\hline
		\small{a}&4&5&\\
		\hline
		$\mathring{g}_{a}$&$\left(\begin{matrix}
		1.03&0\\	0.36&0.97
		\end{matrix}\right)$&$\left(\begin{matrix}
		1&0\\	0&1
		\end{matrix}\right)$&\\
		\hline
	\end{tabular}
	
	\begin{tabular}{|c|c|c|c|c|}
		\hline
		\diagbox{\small{a}}{$|\mathring{\textbf{z}}_{a,b}\rangle$}{\small{b}}&7&8&9&10\\
		\hline
		6&(1,-0.18 - 0.26i)&(1,1)&(1,0.42 + 0.22i)&(1,-0.33 + 0.94i)\\
		\hline
		7&\diagbox{}{}&(1,-1.94 + 1.26i)&(-0.1 - 0.43i)&(1,-0.08 - 0.12i)\\
		\hline
		8&\diagbox{}{}&\diagbox{}{}&(1,0.03 + 1.i)&(1,0.22 - 3.72i)\\
		\hline
		9&\diagbox{}{}&\diagbox{}{}&\diagbox{}{}&(1,-0.13 + 0.74i)\\
		\hline
	\end{tabular}
	
	}
\end{table}

\begin{table}[H]
	\centering\caption{The real critical point $\mathring{g}_{a},\, \mathring{\textbf{z}}_{a,b}$ for the 2nd 4-simplex with points 12456.}\label{tab:ga2}
	\scalebox{0.74}
	{
	\begin{tabular}{|c|c|c|c|}
		\hline
		\small{a}&6&7&8\\
		\hline
		$\mathring{g}_{a}$ &$\left(\begin{matrix}
		1&0\\	0&1
		\end{matrix}\right)$&$\left(\begin{matrix}
		0.82 &0.09-0.13i\\	0.09+0.13i&1.26
		\end{matrix}\right)$&$\left(\begin{matrix}
		0.97&0.34\\	0&1.03
		\end{matrix}\right)$\\
		\hline
		\small{a}&9&10&\\
		\hline
		$\mathring{g}_{a}$ &$\left(\begin{matrix}
		1.04&0.09+0.25i\\	0.09-0.25i&1.04
		\end{matrix}\right)$&$\left(\begin{matrix}
		1.26&0.09-0.13i\\	0.09+0.13i&0.82
		\end{matrix}\right)$&\\
		\hline
	\end{tabular}
	
	\begin{tabular}{|c|c|c|c|c|}
		\hline
		\diagbox{\small{a}}{$|\mathring{\textbf{z}}_{a,b}\rangle$}{\small{b}}&7&8&9&10\\
		\hline
		6&(1,-0.18 - 0.26i)&(1,1)&(1,0.42 + 0.22i)&(1,-0.33 + 0.94i)\\
		\hline
		7&\diagbox{}{}&(1,-1.94 + 1.26i)&(-0.1 - 0.43i)&(1,-0.08 - 0.12i)\\
		\hline
		8&\diagbox{}{}&\diagbox{}{}&(1,0.03 + 1.i)&(1,0.22 - 3.72i)\\
		\hline
		9&\diagbox{}{}&\diagbox{}{}&\diagbox{}{}&(1,-0.13 + 0.74i)\\
		\hline
	\end{tabular}
	}
\end{table}

\begin{table}[H]
	\centering\caption{The real critical point $\mathring{g}_{a},\, \mathring{\textbf{z}}_{a,b}$ for the 3rd 4-simplex with points 13456.}\label{tab:ga3}
	\scalebox{0.6}
	{
	\begin{tabular}{|c|c|c|c|}
		\hline
		\small{a}&11&12&13\\
		\hline
		$\mathring{g}_{a}$ &$\left(\begin{matrix}
		1.04&-0.02\\	-0.36&0.97
		\end{matrix}\right)$&$\left(\begin{matrix}
		0.97 &-0.36\\	0&1.03
		\end{matrix}\right)$&$\left(\begin{matrix}
		1.02+0.001i&-0.19+0.003i\\	-0.19-0.003i&1.01-0.001i
		\end{matrix}\right)$\\
		\hline
		\small{a}&14&15&\\
		\hline
		$\mathring{g}_{a}$ &$\left(\begin{matrix}
		1.012 - 0.001i&-0.19 - 0.006i\\	-0.19 + 0.006i&1.02 + 0.001i
		\end{matrix}\right)$&$\left(\begin{matrix}
		1.01 + 0.001i&-0.19+ 0.003i\\	-0.19 - 0.003i&1.02 - 0.001i
		\end{matrix}\right)$&\\
		\hline
	\end{tabular}
	
	\begin{tabular}{|c|c|c|c|c|c|}
		\hline
		\diagbox{\small{a}}{$|\mathring{z}_{a,b}\rangle$}{\small{b}}&11&12&13&14&15\\
		\hline
		11&\diagbox{}{}&(1,1)&(1,0.1 + 3.73i)&\diagbox{}{}&\diagbox{}{}\\
		\hline
		12&\diagbox{}{}&\diagbox{}{}&(1,1.41 + 0.31i)&(1, 0.92 - 0.4i)&(1,0.68 + 0.15i)\\
		\hline
		13&\diagbox{}{}&\diagbox{}{}&\diagbox{}{}&(1,0.68 + 1.52i)&(1,5.35 + 0.08i)\\
		\hline
		14&(1,0.64 + 0.77i)&\diagbox{}{}&\diagbox{}{}&\diagbox{}{}&(1,0.67 - 1.5i)\\
		\hline
		15&(1,1.92 - 1.16i)&\diagbox{}{}&\diagbox{}{}&\diagbox{}{}&\diagbox{}{}\\
		\hline
	\end{tabular}
	}
\end{table}     

All the boundary data $\mathring{r}=(\mathring{j}_{a,b},\mathring{\xi}_{a,b})$ and the data of the real critical point $(\mathring{j}_h, \mathring{g}_{a}, \mathring{\textbf{z}}_{a,b})$ can be found in the Mathematica notebook in \cite{qudx_Delta3.org}.

We focus on the Regge-like boundary data $r=\{j_b,\xi_{eb}\}$. The Regge-like boundary data determines the geometries of boundary tetrahedra that are glued with the shape-matching and orientation-matching conditions \cite{Kaminski:2017eew} to form the boundary Regge geometry on $\partial\Delta_3$. Then the resulting boundary segment-lengths uniquely determine the 4d Regge geometry $\mathbf{g}(r)$ on $\Delta_3$. The above $\mathring{r}=(\mathring{j}_{a,b},\mathring{\xi}_{a,b})$ is an example of the Regge-like boundary data, which determine the flat geometry $\mathbf{g}(\mathring{r})$ on $\Delta_3$. Generic Regge-like boundary conditions $r$ determines the curved geometries $\mathbf{g}(r)$.

\subsection{Parametrization of variables}\label{parametrization}

Given the Regge-like boundary condition $r$, we find the \emph{pseudo-critical point} $({j}^0_h,{g}^0_{a},{\textbf{z}}^0_{a,b})$ inside the integration domain, where $({j}^0_h,{g}^0_{a},{\textbf{z}}^0_{a,b})$ only satisfies $\re(S)=\partial_{g_{ve}} S=\partial_{\mathbf{z}_{vf}}S=0$ but does not necessarily satisfies $\partial_{j_h}S=4\pi i k_h$. The {pseudo-critical point} $({j}^0_h,{g}^0_{a},{\textbf{z}}^0_{a,b})$ is the critical point of the spinfoam amplitude with fixed $j_h,j_b$ \cite{Han:2011re}, and endows the Regge geometry $\mathbf{g}(r)$ and all $s_v=+1$ to $\Delta_3$. It reduces to the real critical point $(\mathring{j}_h,\mathring{g}_{a},\mathring{\textbf{z}}_{a,b})$ when $r=\mathring{r}$. $({j}^0_h,{g}^0_{a},{\textbf{z}}^0_{a,b})$ is close to $(\mathring{j}_h,\mathring{g}_{a},\mathring{\textbf{z}}_{a,b})$ in the integration domain when $r$ is close to $\mathring{r}$ (by the natural metrics on the integration domain and the space of $r$). The data of the pseudo-critical points are given in \cite{qudx_Delta3.org}.

We consider a neighborhood enclose both $({j}^0_h,{g}^0_{a},{\textbf{z}}^0_{a,b})$ and $(\mathring{j}_h,\mathring{g}_{a},\mathring{\textbf{z}}_{a,b})$. We use the following real parametrizations of the integration variables, according to the gauge-fixing in Section \ref{Real critical Point},
\begin{itemize}
	\item As $a=1, 10, 15$, $g_{a}={g}^0_{a}$. 
	
	\item As $a=5,8,12$, $g_a$ is gauge-fixed to be an upper triangular matrix {(${g}^0_{a}$ is upper triangular)}:
	\be
	g_a&=&{g}^0_{a}\left(\begin{array}{cc}
    1+\frac{x^{1}_{a}}{\sqrt{2}} & \frac{x^2_{a}+i y^2_{a}}{\sqrt{2}} \\
    0 & \mu_{a}
    \end{array}\right),\label{ga1}
	\ee
	here, $\mu_{a}$ is determined by $\det(g_a)=1$.
	\item As $a=2,3,4,6,7,9,11,13,14$, $g_a$ is parameterized as:
	\be\footnotesize
	g_{a}&=&{g}^0_{a}\left(\begin{array}{cc}
    1+\frac{x^{1}_{a}+i y^{1}_{a}}{\sqrt{2}} & \frac{x^{2}_{a}+i y^{2}_{a}}{\sqrt{2}} \\
    \frac{x^{3}_{a}+i y^{3}_{a}}{\sqrt{2}} & \mu_{a}
    \end{array}\right) \label{ga2}
	\ee
	\item The spinors are parametrized by two real parameters: 
    \be
    \textbf{z}_{a,b}&=&(1,{\a}^0_{a,b}+x_{a,b}+iy_{a,b}). \label{zab}
    \ee
    where ${\a}^0_{a,b}$ is the second component of ${\textbf{z}}^0_{a,b}$.
    \item For the internal spin $j_h$, we parametrize it by one real parameter 
    \be
    j_h={j}^0_h+\mathfrak{j}, \quad \mathfrak{j}\in\mathbb{R}\label{jab}
    \ee
\end{itemize}
We denote by $x\in\R^{124}$ these 124 real variables $\fj, x_a^{1,2,3},y_a^{1,2,3},x_{a,b},y_{a,b}$ parametrizing $(j_h,g_{a}, \textbf{z}_{a,b})$. The parametrizations define the coordinate chart covering the neighborhood enclosing both $x_0=({j}^0_h,{g}^0_{a},{\textbf{z}}^0_{a,b})$ and $\mathring{x}=(\mathring{j}_h,\mathring{g}_{a},\mathring{\textbf{z}}_{a,b})$. This neighborhood is large since the parametrizations are valid generically. The pseudo-critical point is ${x}_0=(0,0,...,0)$, which contains 124 zero components. The spinfoam action can be expressed as $S(r,x)$. The integrals in \eqref{integralFormAmp22} (for $\ck=\Delta_3$) can be expressed as
\be
\int \mathrm{d}^{N}x\,\mu(x)\,e^{\lambda S(r,x)},\label{typeint}
\ee
where $N=124$. Both $S(r,x)$ and $\mu(x)$ is analytic in the neighborhood of $\mathring{x}$. We only focus on the integral $k_h=0$ in \eqref{integralFormAmp22}, since other $k_h\neq 0$ integrals has no real critical point by the boundary data $\mathring{r}$. $S(r,x)$ can be analytic continue to a holomorphic function $\cs(r,z)$, $z\in\C^N$ in a complex neighborhood of $\mathring{x}$. Here the analytic continuation is obtained by simply extending $x\in \R^N$ to $z\in \C^N$. The formal discussion of the analytic continuation of the spinfoam action is given in \cite{Han:2021rjo}.

\subsection{Geometrical variations}\label{Geometrical perturbations}

The different regimes of the boundary data $r$ result in different large-$\l$ asymptotic behavior of $A(\Delta_3)$. 
\begin{itemize}
    \item Regime 1: fixing the boundary data $r=\mathring{r}$, Section \ref{Real critical Point} gives numerically the real critical point for the flat geometry $\mathbf{g}(\mathring{r})$, whose deficit angle is $\delta_h=0$. $e^{S(\mathring{r},\mathring{x})}$ evaluated at the real critical points $\mathring{x}$ gives the dominant contribution to the asymptotic amplitude.
    \be
    \int \mathrm{d}^{N}x\,\mu(x)\,e^{\lambda S(\mathring{r},x)}\sim\left(\frac{1}{\lambda}\right)^{\frac{N}{2}} \frac{e^{\lambda S(\mathring{r},\mathring{x})}\mu(\mathring{x})}{\sqrt{\det\left(-\delta^2_{x,x}S(r,\mathring{x})/2\pi\right)}}\lt[1+O(1/\l)\rt]. \label{realappro}
    \ee
    The asymptotics behaves as a power-law in $1/\l$. Here we only focus on the contribution from the single real critical point $\mathring{x}$. There is another real critical point which we discuss in a moment.
  
     \item Regime 2: fixing the boundary data $r$ which determine the segment-lengths for a curved geometry $\mathbf{g}(r)$, the real critical point is absent, then the integral is suppressed faster than any polynomial in $1/\l$:
     \be
      \int \mathrm{d}^{N}x\,\mu(x)\,e^{\lambda S(r,x)}=O(\lambda^{-K}),\quad \forall\,K>0. \label{Apmsupp}
     \ee
\end{itemize}

The above asymptotic behavior is based on fixing $r$ and send $\l$ to be large. However, in order to clarify contributions from curved geometries and compare to the contribution from the flat geometry, we should also let $r$ vary and have an interpolation between two regimes (\ref{realappro}) and (\ref{Apmsupp}). This motivates us to use the complex critical point of the analytic continued action $\cs(r,z)$.

To obtain the curved geometries, we fix the geometries of the 4-simplices $12345$ and $13456$, but change the geometry of 4-simplex $12356$ by varying the length of $l_{26}$ (the length of the line segment connecing point 2 and 6) from $5.44+9.2\times10^{-17}$ to $5.44+9.2\times10^{-5}$ with interval size $10^{-1}$. For each given $l_{26}$, we repeat the steps in Section \ref{delta3} and \ref{Real critical Point} to reconstruct the geometry and compute all the geometric quantities, such as the triangle areas, the 4-d normals of tetrahedra, the 3-d normals of triangles, $\xi_{a,b}$, the deficit angle, etc. Part of the data for the fluctuation $\delta l_{26}=l_{26}-\mathring{l}_{26}$ and the corresponding deficit angle $\delta_{h}$ are shown in Table \ref{tab:delta}. These new geometries $\textbf{g}(r)$ are curved geometries because of non-zero deficit angles. 
\begin{table}[H]
	\centering\caption{Each cell of the table is the value of internal deficit angle $\delta_{h}$ with fluctuation $\delta l_{26}=l_{26}-\mathring{l}_{26}$. }\label{tab:delta}
	\scalebox{0.9}
	{
	\begin{tabular}{|c|c|c|c|c|c|c|c|c|c|c|}
		\hline
		\small{$\delta l_{26}$}&$9.2\times10^{-17}$&$8.3\times10^{-15}$&$7.3\times10^{-14}$&$6.4\times10^{-13}$&$4.6\times10^{-11}$&$8.3\times10^{-10}$&$7.3\times10^{-9}$&$4.6\times10^{-6}$&$9.2\times10^{-6}$&$9.2\times10^{-5}$\\
		\hline
		$\delta_{h}$&$2.0\times10^{-16}$&$1.8\times10^{-14}$&$1.6\times10^{-13}$&$1.40\times10^{-12}$&$1.00\times10^{-10}$&$1.81\times10^{-9}$&$1.61\times10^{-8}$&$1.00\times10^{-5}$&$2.\times10^{-5}$&$0.0002$\\
		\hline
	\end{tabular}
	}
\end{table}

\subsection{Numerical solving complex critical points and error estimate}\label{Numerical solving complex critical points}

At each curved geometry $\textbf{g}(r)$, the real critical point is absent for $\delta_h\neq0$. We numerically compute the complex critical point $Z(r)$ satisfying the complex critical equations $\partial_z\cs(r,z)=0$ with Newton-like recursive procedure. First, we linearize $\partial_z\cs(r,z)=0$ at the pseudo-critical point $x_0\in\R^{124}$. Then, we have the linear system of equations
\be
\partial^2_{z,z}\cs({r},{x}_0)\cdot\delta z_{1}+\partial_{z}\cs(r,{x}_0)\simeq 0, \label{eq1}
\ee
We obtain $z_1=x_0+\delta z_1$ by the solution $\delta z_1$. We again linearize $\partial_z\cs(r,z)=0$ at $z_1$, 
\be
\partial^2_{z,z}\cs({r},{z}_1)\cdot\delta z_{2}+\partial_{z}\cs(r,{z}_1)\simeq 0, \label{eq1}
\ee
we obtain $z_2=z_1+\delta z_2$ by the solution $\delta z_2$. We iterate and linearize the complex critical equations at $z_2,z_3,\cdots,z_{n-1}$. The resulting $z_n=z_{n-1}+\delta z_n$ should approximates the complex critical points $Z(r)$ arbitrarily well for sufficiently large $n$. In practise, $n=4$ turns out to be sufficient for our calculation.  The numerical results of complex critical point for each geometry $r$ can be found in Mathematica notebook \cite{qudx_Delta3.org}.


The absolute error of numerically solving $\partial_z\cs(r,z)=0$ is measured by
\be
\varepsilon=\max\left|\partial_z \cs(r,z_n)\right|.
\ee
We have $z_n$ well-approximate the complex critical point $Z(r)$ if $\varepsilon$ is small. $\varepsilon$ in the case of $\gamma=0.1,\,n=4$ for some deficit angles are shown in Table \ref{tab:ordern}. The absolute errors is small and scales as $\varepsilon\approx 1.31 \delta_h^5$ at $n=4$. 
\begin{table}[H]
        \centering\caption{Deficit angles $\delta_h$ and corresponding absolute errors}\label{tab:ordern}
        \scalebox{0.9}
        {
        \begin{tabular}{|c|c|c|c|c|c|c|c|c|c|c|c|}
        \hline
        $\delta_h$ & $2\times 10^{-16}$ & $1.8\times 10^{-14}$&$1.6\times10^{-13}$&$1.4\times10^{-12}$ &$1.0\times 10^{-10}$&$1.8\times10^{-9}$&$1.6\times10^{-8}$&$1.6\times10^{-5}$&$2\times10^{-5}$&0.0002\\
        \hline
        $\varepsilon$ & $4.3 \times 10^{-79}$ & $2.5\times 10^{-69}$&$1.4\times10^{-64}$&$7.1\times10^{-60}$&$1.3\times10^{-50}$&$2.5\times10^{-44}$&$1.4\times10^{-39}$&$1.4\times10^{-24}$&$4.2\times10^{-24}$&$4.2\times10^{-19}$ \\
        \hline
        \end{tabular}
        }
    \end{table}

\subsection{Flipping orientations and numerical results}\label{NumericalResult}


So far we only consider one real critical point with all $s_v=+1$. When we take into account different orientations of each 4-simplex $v$, there is another real critical point with all $s_v=-1$. Other 6 discontinuous orientations violate the flatness constraint $\g {\delta}^{s}_h=\g\sum_vs_v\Theta_h(v)=0$ thus do not correspond to any real critical point. Given $\mathring{r}$, Table \ref{dresseddefcit} lists $\delta^s_h$'s at different orientations.

\begin{table}[H]
    \centering\caption{}
    \begin{tabular}{|c|c|c|c|c|c|c|c|c|}
    \hline
          $s$ & $+++$ & $---$ & $++-$ & $--+$& $+--$&$-++$&$-+-$&$+-+$\\
         \hline
         $\delta^s_h$ & 0 &0 & 0.043 & $-0.043$ & 0.72 & $-0.72$ & $-0.68$ & 0.68\\
         \hline
    \end{tabular}
    \label{dresseddefcit}
\end{table}

As in Section \ref{Geometrical perturbations}, we deforming the boundary data $r=\mathring{r}+\delta r$ to obtain curved geometries. Both real critical points with all $s_v=+$ and all $s_v=-$ move smoothly away from the real plane, and become complex critical points. We numerically compute the other complex critical point $Z'(r)$ with all $s_v=-$ by the same procedure as in Section \ref{Numerical solving complex critical points}. We compute 
\be
\delta \ci(r)=\cs(r,Z(r))-i\ci_R[\textbf{g}(r)],\quad \delta \ci'(r)=\cs(r,Z'(r))+i\ci_R[\textbf{g}(r)]
\ee
for the sequences of $r$ of curved geometries. $\delta\ci$ and $\delta\ci'$ associate to two continuous orientations $s_v=+$ and $s_v=-$ respectively. Part of the results are shown in Table \ref{tab:diffvstheta} at $\gamma=0.1$. 
\begin{table}[H]
	\centering\caption{$\delta \ci(r)$ and $\delta \ci'(r)$ at different deficit angles $|\delta^{s_v}_{h}|$.}\label{tab:diffvstheta}
	\small
	\scalebox{0.8}
	{
	\begin{tabular}{|c|c|c|c|c|c|c|}
		\hline
		$|\delta^{s_v}_{h}|$&$2.\times10^{-15}$&$1.4\times10^{-12}$&$1\times10^{-10}$&$1.61\times10^{-8}$&$2\times10^{-4}$\\
		\hline
		$\delta\ci$&$-6.36\times10^{-34}-3.34\times10^{-35}$&$-3.12\times10^{-28}-1.63\times10^{-27}i$&$-1.59\times10^{-24}-8.34\times10^{-24}i$&$-4.07\times10^{-20}-2.13\times10^{-19}i$&$-6.30\times10^{-12}-3.32\times10^{-11}i$\\
		\hline
		$\delta\ci'$&$-6.36\times10^{-34}+3.34\times10^{-35}$&$-3.12\times10^{-28}+1.63\times10^{-27}i$&$-1.59\times10^{-24}+8.34\times10^{-24}i$&$-4.07\times10^{-20}+2.13\times10^{-19}i$&$-6.30\times10^{-12}+3.32\times10^{-11}i$\\
		\hline
		
	\end{tabular}
	}
\end{table}

The best-fit functions are 
\be
\delta \ci(r) = a_2 (\delta^+_{h})^2 + a_3 (\delta^+_{h})^3+a_4 (\delta^+_{h})^4+ O((\delta^+_{h})^5),\label{expression1}\\
\delta \ci'(r) = {a}^*_2 (\delta^-_{h})^2 - {a}^*_3 (\delta^-_{h})^3+{a}^*_4 (\delta^-_{h})^4+ O((\delta^-_{h})^5),\label{expression2}
\ee 
where $\delta_h^\pm\equiv \delta_h^{\pm\pm\pm}$. ${a}^*_i$ is the complex conjugate of $a_i$. The best fit coefficient $a_i$ and the corresponding fitting errors are
\be
a_2&=&-0.00016_{\pm10^{-17}}-0.00083_{\pm10^{-16}}i,\nonumber\\
a_3&=&-0.0071_{\pm10^{-13}}-0.011_{\pm10^{-12}}i,\nonumber\\
a_4&=&-0.059_{\pm10^{-9}}+0.070_{\pm10^{-8}}i,
\ee
Fig \ref{thetaS}(a) demonstrates the excellent agreement between the numerical data and the fitted polynomial function at $\gamma=0.1$ and $\lambda=10^{11}$. 

Then, the asymptotic amplitude is obtained
\be
A(\Delta_3)&=&\lt(\frac{1}{\l}\rt)^{60}\lt[\sn_r^+ e^{i\l\ci_R[\mathbf{g}(r)]+\l\delta \ci(r)}+\sn_r^- e^{-i\l\ci_R[\mathbf{g}(r)]+\l\delta \ci'(r)}\rt]\lt[1+O(1/\l)\rt].
\ee 
At $\g=0.1,\delta^{\pm}_h\simeq\pm 2\times10^{-4}$, we have $\sn_r^+/\sn_r^-\simeq 0.001+0.005 i$, $\ci_R\simeq-0.22\g$, $\delta\ci^+\simeq-6.30\times10^{-12}-3.32\times10^{-11}i$ and $\delta \ci^-\simeq-6.30\times 10^{-12} + 3.32\times10^{-11} i$.


Other 6 discontinuous orientations do not contribute to \eqref{asymp} in the range of $\delta r$ for sufficiently large $\l$, since their $|{\delta}^{s}_h(r)|> 0.01$ are not sufficiently small (see FIG.\ref{delta8}). 


\begin{figure}[H]
\centering
\includegraphics[width=0.5\textwidth]{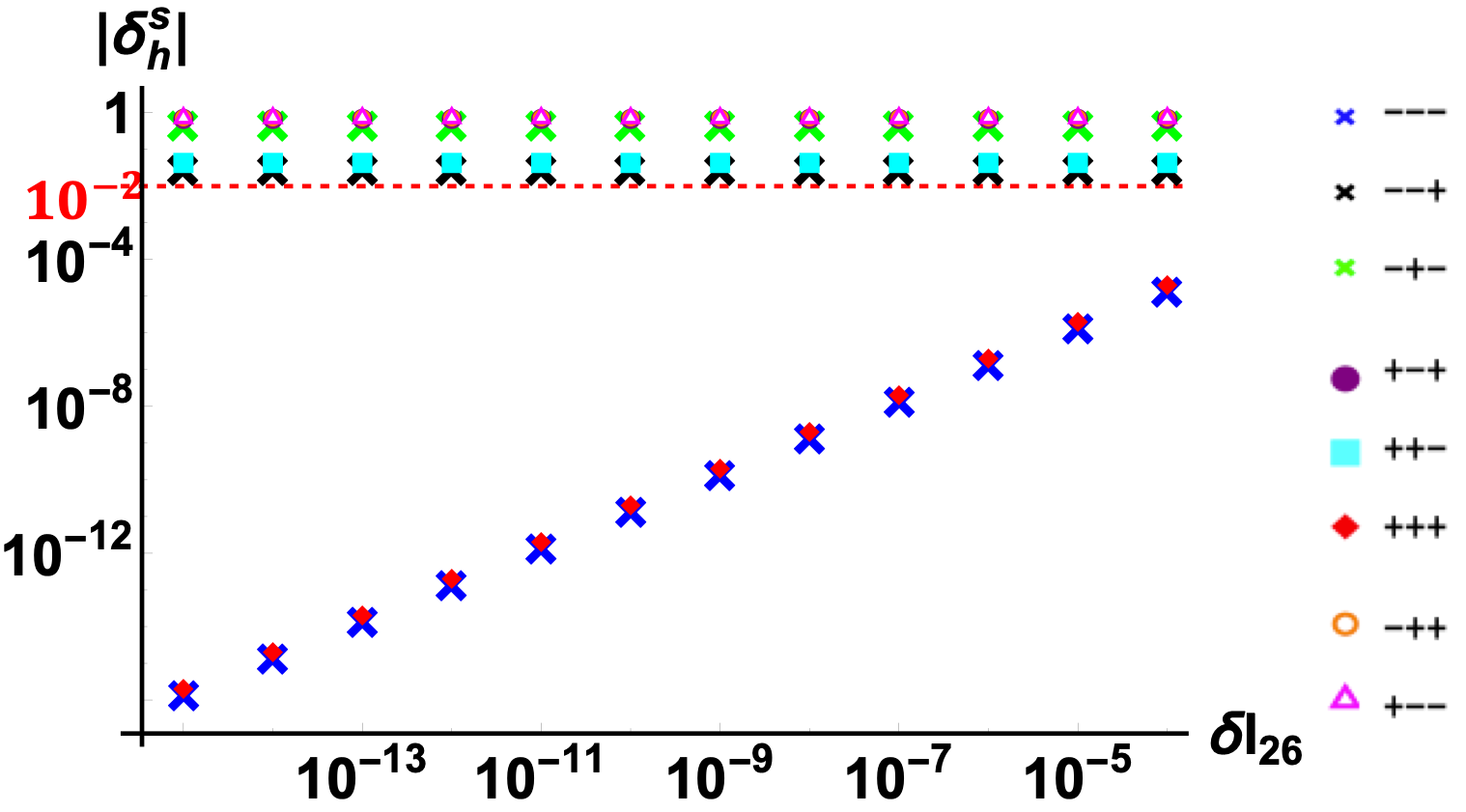}
\caption{The Log-Log plot of $|\delta^s_h|$ for different $s=\{s_v\}_v$ when varying $l_{26}=\mathring{l}_{26}+\delta l_{26}$.}
\label{delta8}
\end{figure}

\section{1-5 Pachner move and ${A(\sig_{\text{1-5}})}$}\label{Pachner moves in a 4d Lorentzian Spinfoam model}

\subsection{Flat geometry, boundary data, and real critial point}

The triangulation $\sig_{\text{1-5}}$ of the 1-5 pachner move is made by five 4-simplices. $\sig_{\text{1-5}}$ is obtained by adding an point $6$ inside a 4-simplex and connecting $6$ to other 5 points of the 4-simplex by 5 line segments $(1,6),(2,6),\cdots,(5,6)$. See Fig.\ref{pic}(b). The dual cable diagram of $\sig_{\text{1-5}}$ is in Fig.\ref{4complex}(b) \interfootnotelinepenalty=10000\footnote{For convenience, the indexes of group variables in FIG. \ref{4complex}(b) are $a=1,2,\cdots,25$, the corresponding tetrahedra $e$ are labeled by the numbers in FIG.\ref{pic}(b). The correspondence are: $g_1\rightarrow e_{1,2,3,4}$, $g_2\rightarrow e_{1,2,3,6}$, $g_3\rightarrow e_{1,2,4,6}$, $g_4\rightarrow e_{1,3,4,6}$, $g_5\rightarrow e_{2,3,4,6}$, $g_6\rightarrow e_{1,2,3,5}$, $g_7\rightarrow e_{1,2,3,6}$, $g_8\rightarrow e_{1,2,5,6}$, $g_9\rightarrow e_{1,3,5,6}$, $g_{10}\rightarrow e_{2,3,5,6}$, $g_{11}\rightarrow e_{1,2,4,5}$, $g_{12}\rightarrow e_{1,2,4,6}$, $g_{13}\rightarrow e_{1,2,5,6}$, $g_{14}\rightarrow e_{1,4,5,6}$, $g_{15}\rightarrow e_{2,4,5,6}$, $g_{16}\rightarrow e_{1,3,4,5}$, $g_{17}\rightarrow e_{1,3,4,6}$, $g_{18}\rightarrow e_{1,3,5,6}$, $g_{19}\rightarrow e_{1,4,5,6}$, $g_{20}\rightarrow e_{3,4,5,6}$, $g_{21}\rightarrow e_{2,3,4,5}$, $g_{22}\rightarrow e_{2,3,4,6}$, $g_{23}\rightarrow e_{2,3,5,6}$, $g_{24}\rightarrow e_{2,4,5,6}$, $g_{25}\rightarrow e_{3,4,5,6}$.} (see also \cite{Banburski:2014cwa}). $\sig_{\text{1-5}}$ consists of 10 boundary triangles $b$ (dual to black strands in Fig. \ref{4complex}(b)) and 10 internal triangles $h$ (dual to colored loops in Fig. \ref{4complex}(b)). Here, we set the coordinates of $P_1,P_2,P_3,P_4,P_5$ the same as Eq.(\ref{P6}). The coordinate of the point 6 is 
\be
P_6=\left(-0.068, -0.27, -0.50, -1.30\right),
\ee
$P_1,\cdots,P_6$ determines a flat Regge geometry on $\sig_{\text{1-5}}$. We obtain five Lorentzian 4-simplices, $S_{12346},S_{12356},S_{12456},S_{13456},S_{23456}$ with all tetrahedra and triangles space-like. The lengths of the internal line segments are $l_{16}\approx 2.01,\, l_{26}\approx 6.66,\,l_{36}\approx 4.72, \,l_{46}\approx 0.54,\,l_{56}\approx 6.19$. The 4-d normals are determined by Eq.(\ref{4dnormals}). For convenience, we choose $(N_a)_\mu$ with $a=2,6,13,18,23$ to be $(-1,0,0,0)$ as reference for each 4-simplex. Hence, the 4-d normals $(N_a)_\mu$ in each 4-simplex are given by: 
\begin{itemize}
    \item The first 4-simplex 12346:
    \be
   & N_1=(1.02,-0.06,0.17,0),\quad N_2=(-1,0,0,0),\quad N_3=(-1.15,0.07,-0.53,0.19),\nonumber\\
    &N_4=(1.50,0.98,-0.54,0),\quad N_5=(-1.04,0.06,-0.28,-0.06).\nonumber
    \ee
    \item The second 4-simplex 12356:
    \be
    & N_6=(-1,0,0,0),\quad N_7=(1.02,-0.06,0.17,0),\quad N_8=(1.00,-0.03,-0.04,0.07),\nonumber\\
    &N_9=(1.03,0.26,0,0),\quad N_{10}=(1.00,-0.02,-0.02,-0.04).\nonumber
    \ee
    \item The third 4-simplex 12456:
    \be
    & N_{11}=(1.0,-0.091,-0.13,0.22),\quad N_{12}=(1.3,-0.11,0.79,-0.28),\quad N_{13}=(-1,0,0,0),\nonumber\\
    &N_{14}=(1.1, 0.50, 0.077,-0.13),\quad N_{15}=(-1.5,0.14,0.19,-1.1).\nonumber
    \ee
    \item The fourth 4-simplex 13456:
    \be
    & N_{16}=(1.0,0.10,0,0),\quad N_{17}=(-1.2,-0.57,0.30,0),\quad N_{18}=(-1,0,0,0),\nonumber\\
    &N_{19}=(-1.0,-0.19,-0.029,0.049),\quad N_{20}=(-1.0,-0.14,-0.012,-0.020).\nonumber
    \ee
    \item The fifth 4-simplex 23456:
    \be
    & N_{21}=(1.0,-0.11,-0.15,-0.26),\quad N_{22}=(1.1, -0.11, 0.49,0.10),\quad N_{23}=(-1,0,0,0),\nonumber\\
    &N_{24}=(1.6,-0.16,-0.22,1.3),\quad N_{25}=(1.1,0.42,0.037,0.064).\nonumber
    \ee
\end{itemize}
Then we compute all dihedral angles $\theta_{a,b}$ in each 4-simplex. We check that all deficit angles $\delta_h,\,h=1,2,\cdots,10$ hinged by 10 internal triangles vanish
\be
0=\delta_1&=&\theta_{2,3}+\theta_{12,13}+\theta_{8,7}\approx -0.54+0.77-0.23,\qquad0=\delta_2=\theta_{2,4}+\theta_{17,18}+\theta_{9,7}\approx 0.965-0.604-0.361,\nonumber\\
0=\delta_3&=&\theta_{3,4}+\theta_{17,19}+\theta_{14,12}\approx 1.37-0.47-0.90,\qquad0=\delta_4=\theta_{8,9}+\theta_{18,19}+\theta_{14,13}\approx -0.3-0.2+0.5,\nonumber\\
0=\delta_5&=&\theta_{2,5}+\theta_{22,23}+\theta_{10,7}\approx -0.29+0.49-0.2,\qquad\, 0=\delta_6=\theta_{3,5}+\theta_{22,24}+\theta_{15,12}\approx -0.3-1.2+1.5,\nonumber\\
0=\delta_7&=&\theta_{8,10}+\theta_{23,24}+\theta_{15,13}\approx -0.12+1.07-0.95,\quad0=\delta_8=\theta_{4,5}+\theta_{22,25}+\theta_{20,17}\approx 1.18-0.69-0.49,\nonumber\\
0=\delta_9&=&\theta_{9,10}+\theta_{23,25}+\theta_{20,18}\approx -0.28+0.42-0.14,\quad0=\delta_{10}=\theta_{14,15}+\theta_{24,25}+\theta_{20,19}\approx 1.26-1.17-0.09.\nonumber
\ee

We adapt the similar steps as in $\Delta_3$ with Eq.(\ref{O13}), (\ref{nab}) and (\ref{xi}) to compute the normalized spinors $\xi_{a,b}$. We compute areas $j_{a,b}$ in each 4-simplex: 
\begin{table}[H]
	\scalebox{0.74}
	{
	\begin{tabular}{|c|c|c|c|c|c|}
		\hline
		\diagbox{\small{a}}{$\mathring{\xi}_{a,b}$}{\small{b}}&1&2&3&4&5\\
		\hline
		1&\diagbox{}{}&(0.71,-0.24+0.67i)&(0.86,0.01-0.51i)&(0.71,0.57-0.43i)&(0.51,0.02-0.86i)\\
		\hline
		2&(0.64,-0.26+0.72i)&\diagbox{}{}&(0.51+0.02i,-0.13+0.85i)&(0.66-0.04i,-0.64+0.40i)&(0.71+0.01i,-0.16+0.68i)\\
		\hline
		3&(0.97-0.03i,-0.03-0.25i)&(0.32,-0.14+0.94i)&\diagbox{}{}&(0.42-0.01i,-0.49+0.76i)&(-0.99-0.02i,-0.035-0.11i)\\
		\hline
		4&(0.56-0.02i,0.67-0.49i)&(0.82+0.05i,-0.51+0.24i)&(0.80+0.01i,-0.44+0.40i)&\diagbox{}{}&(0.54-0.02i,0.65-0.53i)\\
		\hline
		5&(0.69-0.05i,-0.01-0.72i)&(0.61,-0.19+0.77i)&(0.99-0.02i,-0.05-0.15i)&(0.81+0.09i,0.48-0.33i)&\diagbox{}{}\\
		\hline
	\end{tabular}	

	\begin{tabular}{|c|c|c|c|c|c|}
		\hline
		\diagbox{\small{a}}{$\mathring{j}_{a,b}$}{\small{b}}&1&2&3&4&5\\
		\hline
		1&\diagbox{}{}&5&2&2&2\\
		\hline
		2&\diagbox{}{}&\diagbox{}{}&1.7&0.96&2.8\\
		\hline
		3&\diagbox{}{}&\diagbox{}{}&\diagbox{}{}&0.29&0.60\\
		\hline
		4&\diagbox{}{}&\diagbox{}{}&\diagbox{}{}&\diagbox{}{}&0.76\\
		\hline
	\end{tabular}	
	}
\end{table}

\begin{table}[H]
	\scalebox{0.77}
	{
	\begin{tabular}{|c|c|c|c|c|c|}
		\hline
		\diagbox{\small{a}}{$\mathring{\xi}_{a,b}$}{\small{b}}&6&7&8&9&10\\
		\hline
		6&\diagbox{}{}&(0.71,-0.24 + 0.67i)&(0.30,0.55+0.78i)&(0.71,-0.71)&(0.95,0.17+0.25i)\\
		\hline
		7&(0.64,-0.26+0.72i)&\diagbox{}{}&(0.51+0.01i,-0.13+0.85i)&(0.66-0.04i,-0.64+0.40i)&(0.71+0.01i,-0.16+0.68i)\\
		\hline
		8&(0.33,0.55+0.77i)&(0.59+0.02i,-0.12+0.80i)&\diagbox{}{}&(0.62-0.02i,0.77+0.11i)&(0.14,0.57+0.81i)\\
		\hline
		9&(0.79,-0.61)&(0.78-0.04i,-0.53+0.32i)&(0.51-0.01i,0.85+0.12i)&\diagbox{}{}&(0.75,0.66-0.06i)\\
		\hline
		10&(0.96,0.17+0.24i)&(0.78,-0.15+0.61i)&(0.12,0.57+0.81i)&(0.65,-0.76-0.07i)&\diagbox{}{}\\
		\hline
	\end{tabular}	

	\begin{tabular}{|c|c|c|c|c|c|}
		\hline
		\diagbox{\small{a}}{$\mathring{j}_{a,b}$}{\small{b}}&6&7&8&9&10\\
		\hline
		6&\diagbox{}{}&\diagbox{}{}&5&5&5\\
		\hline
		7&5&\diagbox{}{}&\diagbox{}{}&\diagbox{}{}&\diagbox{}{}\\
		\hline
		8&\diagbox{}{}&1.7&\diagbox{}{}&1.6&3.2\\
		\hline
		9&\diagbox{}{}&0.96&\diagbox{}{}&\diagbox{}{}&2.7\\
		\hline
		10&\diagbox{}{}&2.8&\diagbox{}{}&\diagbox{}{}&\diagbox{}{}\\
		\hline
	\end{tabular}	
	}
\end{table}

\begin{table}[H]
	\scalebox{0.72}
	{
	\begin{tabular}{|c|c|c|c|c|c|}
		\hline
		\diagbox{\small{a}}{$\mathring{\xi}_{a,b}$}{\small{b}}&11&12&13&14&15\\
		\hline
		11&\diagbox{}{}&(0.87,-0.01-0.49i)&(0.30,0.55+0.78i)&(0.49,0.82+0.31i)&(0.015,0.58+0.82i)\\
		\hline
		12&(0.97-0.03i,-0.03-0.25i)&\diagbox{}{}&(0.32,-0.14+0.94i)&(0.42-0.01i,-0.48+0.76i)&(0.99-0.02i,-0.036-0.106i)\\
		\hline
		13&(0.33,0.55+0.77i)&(0.59+0.02i,-0.12+0.80i)&\diagbox{}{}&(0.62-0.02i,0.77+0.11i)&(0.14,0.57+0.81i)\\
		\hline
		14&(0.30-0.02i,0.91+0.30i)&(0.75-0.14i,-0.38+0.52i)&(0.41+0.01i,0.90+0.15i)&\diagbox{}{}&(0.09-0.024i,0.94+0.32i)\\
		\hline
		15&(0.14,0.57+0.81i)&(0.94-0.01i,-0.08-0.34i)&(0.21,0.56+0.80i)&(0.32-0.05i,0.86+0.39i)&\diagbox{}{}\\
		\hline
	\end{tabular}	

	\begin{tabular}{|c|c|c|c|c|c|}
		\hline
		\diagbox{\small{a}}{$\mathring{j}_{a,b}$}{\small{b}}&11&12&13&14&15\\
		\hline
		11&\diagbox{}{}&\diagbox{}{}&\diagbox{}{}&2&2\\
		\hline
		12&2&\diagbox{}{}&1.7&\diagbox{}{}&\diagbox{}{}\\
		\hline
		13&5&\diagbox{}{}&\diagbox{}{}&\diagbox{}{}&\diagbox{}{}\\
		\hline
		14&\diagbox{}{}&0.29&1.6&\diagbox{}{}&0.68\\
		\hline
		15&2&\diagbox{}{}&\diagbox{}{}&0.68&\diagbox{}{}\\
		\hline
	\end{tabular}	
	}
\end{table}

\begin{table}[H]
	\scalebox{0.74}
	{
	\begin{tabular}{|c|c|c|c|c|c|}
		\hline
		\diagbox{\small{a}}{$\mathring{\xi}_{a,b}$}{\small{b}}&16&17&18&19&20\\
		\hline
		16&\diagbox{}{}&(0.71,0.59-0.39i)&(0.71,-0.71)&(0.51,0.82+0.27i)&(0.51,-0.82-0.27i)\\
		\hline
		17&(0.56-0.02i,0.67-0.48i)&\diagbox{}{}&(0.82+0.06i,-0.51+0.23i)&(0.80+0.02i,-0.45+0.40i)&(0.54-0.01i,0.66-0.52i)\\
		\hline
		18&(0.79,-0.61)&(0.78-0.04i,-0.53+0.32i)&\diagbox{}{}&(0.51-0.01i,0.85+0.12i)&(0.75,-0.66-0.06i)\\
		\hline
		19&(0.30-0.02i,0.91+0.30i)&(0.75-0.15i,-0.38+0.53i)&(0.41+0.01i,0.90+0.15i)&\diagbox{}{}&(0.1-0.03i,0.95+0.32i)\\
		\hline
		20&(0.46-0.02i,0.85-0.27i)&(0.73-0.02i,0.53-0.43i)&(0.61,-0.79-0.06i)&(0.37+0.02i,0.88+0.30i)&\diagbox{}{}\\
		\hline
	\end{tabular}	

	\begin{tabular}{|c|c|c|c|c|c|}
		\hline
		\diagbox{\small{a}}{$\mathring{j}_{a,b}$}{\small{b}}&16&17&18&19&20\\
		\hline
		16&\diagbox{}{}&\diagbox{}{}&\diagbox{}{}&\diagbox{}{}&2\\
		\hline
		17&2&\diagbox{}{}&0.96&0.29&\diagbox{}{}\\
		\hline
		18&5&\diagbox{}{}&\diagbox{}{}&1.6&\diagbox{}{}\\
		\hline
		19&2&\diagbox{}{}&\diagbox{}{}&\diagbox{}{}&\diagbox{}{}\\
		\hline
		20&\diagbox{}{}&0.76&2.7&0.68&\diagbox{}{}\\
		\hline
	\end{tabular}	
	}
\end{table}

\begin{table}[H]
	\scalebox{0.78}
	{
	\begin{tabular}{|c|c|c|c|c|c|}
		\hline
		\diagbox{\small{a}}{$\mathring{\xi}_{a,b}$}{\small{b}}&21&22&23&24&25\\
		\hline
		21&\diagbox{}{}&(0.49,-0.02-0.87i)&(0.95,0.17+0.25i)&(0.015,-0.58-0.82i)&(0.49,-0.82-0.31i)\\
		\hline
		22&(0.69-0.05i,-0.01-0.72i)&\diagbox{}{}&(0.61,-0.18+0.77i)&(0.99-0.02i,-0.05-0.15i)&(0.81+0.09i,0.48-0.33i)\\
		\hline
		23&(0.96,0.17+0.24i)&(0.78,-0.15+0.61i)&\diagbox{}{}&(0.12,0.57+0.81i)&(0.65,-0.76-0.07i)\\
		\hline
		24&(0.141,0.57+0.81i)&(0.94-0.01i,-0.08-0.34i)&(0.21,0.56+0.80i)&\diagbox{}{}&(0.32-0.05i,0.86+0.39i)\\
		\hline
		25&(0.46-0.02i,0.84-0.27i)&(0.73-0.02i,0.53-0.43i)&(0.61,-0.79-0.06i)&(0.37+0.02i,0.88+0.30i)&\diagbox{}{}\\
		\hline
	\end{tabular}	

	\begin{tabular}{|c|c|c|c|c|c|}
		\hline
		\diagbox{\small{a}}{$\mathring{j}_{a,b}$}{\small{b}}&21&22&23&24&25\\
		\hline
		22&2&\diagbox{}{}&2.8&0.60&0.76\\
		\hline
		23&5&\diagbox{}{}&\diagbox{}{}&3.2&2.7\\
		\hline
		24&2&\diagbox{}{}&\diagbox{}{}&\diagbox{}{}&0.68\\
		\hline
		25&2&\diagbox{}{}&\diagbox{}{}&\diagbox{}{}&\diagbox{}{}\\
		\hline
	\end{tabular}	
	}
\end{table}
The boundary data $\mathring{r}=\{\mathring{j}_{b},\mathring{\xi}_{eb}\}$ are given in the above tables. The real critical point $(\mathring{j}_h,\mathring{g}_{a},\mathring{z}_{a,b})$ corresponding to the above flat Regge geometry are obtained by solving critical point equations Eqs. (\ref{eom1}) and \eqref{eom2}. To remove the gauge freedom, We choose $g_a$, $a=1,6,11,16,21$, to be identity and $g_a$, $a=2,3,8,9,14,15,17,20,22,23$, to be upper triangular matrix. In each 4-simplex, we choose $a=1,6,11,16,21$ as the references and use Eq.(\ref{group}) to obtain critical points $\mathring{g}_a$. The resulting $\mathring{g}_a$ and $\mathring{z}_{a,b}$ are given below. $\mathring{j}_h$ are given by $\mathring{j}_{a,b}$'s of the internal triangles in the above tables. {The critical point endows the continuous orientation $s_v=-1$ to all 4-simplices.}

\begin{table}[H]
	\scalebox{0.7}
	{
	\begin{tabular}{|c|c|c|c|}
		\hline
		\small{a}&1&2&3\\
		\hline
		$\mathring{g}_{a}$ &$\left(\begin{matrix}
		1.02&-0.06-0.17i\\	-0.06+0.17i&1.02
		\end{matrix}\right)$&$\left(\begin{matrix}
		0.99 &-0.06-0.17i\\	0&1.01
		\end{matrix}\right)$&$\left(\begin{matrix}
		0.83&-0.12-0.61i\\	0&1.20
		\end{matrix}\right)$\\
		\hline
		\small{a}&4&5&\\
		\hline
		$\mathring{g}_{a}$ &$\left(\begin{matrix}
		0.99&0.55+0.29i\\	0.25&1.14+0.074i
		\end{matrix}\right)$&$\left(\begin{matrix}
		0.94&-0.12-0.45i\\	0&1.02
		\end{matrix}\right)$&\\
		\hline
	\end{tabular}
	
	\begin{tabular}{|c|c|c|c|c|}
		\hline
		\diagbox{\small{a}}{$|\mathring{z}_{a,b}\rangle$}{\small{b}}&2&3&4&5\\
		\hline
		1&(1,-0.33 + 0.94 i)&(1,0.08 - 0.69 i)&(0.68 - 0.73i)&(1,0.18 - 1.43 i)\\
		\hline
		2&\diagbox{}{}&(1,-0.14 + 1.50 i)&(1,-0.93 + 0.37i)&(1,-0.16 + 0.77i)\\
		\hline
		3&\diagbox{}{}&\diagbox{}{}&(1,-0.93 + 0.48i)&(1,0.078 - 0.58 i)\\
		\hline
		4&\diagbox{}{}&\diagbox{}{}&\diagbox{}{}&(1,0.64 - 0.88i)\\
		\hline
	\end{tabular}
	}
    \end{table}

\begin{table}[H]
	\scalebox{0.72}
	{
	\begin{tabular}{|c|c|c|c|}
		\hline
		\small{a}&6&7&8\\
		\hline
		$\mathring{g}_{a}$ &$\left(\begin{matrix}
		1&0\\	0&1
		\end{matrix}\right)$&$\left(\begin{matrix}
		0.99 &-0.06-0.17i\\	0&1.01
		\end{matrix}\right)$&$\left(\begin{matrix}
		1.03&-0.03+0.045i\\	0&0.96
		\end{matrix}\right)$\\
		\hline
		\small{a}&9&10&\\
		\hline
		$\mathring{g}_{a}$ &$\left(\begin{matrix}
		0.98&0.25\\	0&1.02
		\end{matrix}\right)$&$\left(\begin{matrix}
		0.98&-0.02+0.02i\\	0&1.02
		\end{matrix}\right)$&\\
		\hline
	\end{tabular}
	
	\begin{tabular}{|c|c|c|c|c|c|}
		\hline
		\diagbox{\small{a}}{$|\mathring{z}_{a,b}\rangle$}{\small{b}}&6&7&8&9&10\\
		\hline
		6&\diagbox{}{}&\diagbox{}{}&(1,1.82 + 2.57i)&(1,-1)&(0.18 + 0.26i)\\
		\hline
		7&(1,-0.33 + 0.94i)&\diagbox{}{}&\diagbox{}{}&\diagbox{}{}&\diagbox{}{}\\
		\hline
		8&\diagbox{}{}&(1,-0.14+ 1.50i)&\diagbox{}{}&(1,1.36 + 0.27i)&(1,4.60 + 6.50i)\\
		\hline
		9&\diagbox{}{}&(1,-0.93 + 0.37i)&\diagbox{}{}&\diagbox{}{}&(1,-1.11 - 0.072i)\\
		\hline
		10&\diagbox{}{}&(1,-0.16 + 0.77i)&\diagbox{}{}&\diagbox{}{}&\diagbox{}{}\\
		\hline
	\end{tabular}
	}
    \end{table}
    
    \begin{table}[H]
	\scalebox{0.63}
	{
	\begin{tabular}{|c|c|c|c|}
		\hline
		\small{a}&11&12&13\\
		\hline
		$\mathring{g}_{a}$ &$\left(\begin{matrix}
		1.08&-0.03+0.04i\\	-0.03-0.04i&0.93
		\end{matrix}\right)$&$\left(\begin{matrix}
		0.77 &-0.08-0.62i\\	0.02+0.04i&1.32-0.02i
		\end{matrix}\right)$&$\left(\begin{matrix}
		0.96&0\\	0.03+0.04i&1.04
		\end{matrix}\right)$\\
		\hline
		\small{a}&14&15&\\
		\hline
		$\mathring{g}_{a}$ &$\left(\begin{matrix}
		0.85&0.45-0.11i\\	0&1.18
		\end{matrix}\right)$&$\left(\begin{matrix}
		1.52&-0.14+0.2i\\	0&0.66
		\end{matrix}\right)$&\\
		\hline
	\end{tabular}
	
	\begin{tabular}{|c|c|c|c|c|c|}
		\hline
		\diagbox{\small{a}}{$|\mathring{z}_{a,b}\rangle$}{\small{b}}&11&12&13&14&15\\
		\hline
		11&\diagbox{}{}&\diagbox{}{}&\diagbox{}{}&(1,1.77 + 0.80i)&(1,9.6 + 13.58i)\\
		\hline
		12&(1,0.03 - 0.62i)&\diagbox{}{}&(1,-0.23 + 1.31i)&\diagbox{}{}&\diagbox{}{}\\
		\hline
		13&(1,1.82 + 2.57i)&\diagbox{}{}&\diagbox{}{}&\diagbox{}{}&\diagbox{}{}\\
		\hline
		14&\diagbox{}{}&(1,-0.84 + 0.33i)&(1,1.21 + 0.14i)&\diagbox{}{}&(1,5.92 + 4.04i)\\
		\hline
		15&\diagbox{}{}&(1,0.027 - 0.53i)&(1,6.48 + 9.17i)&\diagbox{}{}&\diagbox{}{}\\
		\hline
	\end{tabular}
	}
    \end{table}
    
    \begin{table}[H]
	\scalebox{0.7}
	{
	\begin{tabular}{|c|c|c|c|}
		\hline
		\small{a}&16&17&18\\
		\hline
		$\mathring{g}_{a}$ &$\left(\begin{matrix}
		1.00&-0.07\\	-0.07&1.00
		\end{matrix}\right)$&$\left(\begin{matrix}
		0.96 &0.27+0.28i\\	0&1.04
		\end{matrix}\right)$&$\left(\begin{matrix}
		1.02&0\\	-0.26&0.98
		\end{matrix}\right)$\\
		\hline
		\small{a}&19&20&\\
		\hline
		$\mathring{g}_{a}$ &$\left(\begin{matrix}
		0.96+0.01i&0.19-0.06i\\	-0.26-0.38i&0.99
		\end{matrix}\right)$&$\left(\begin{matrix}
		1.01&-0.12-0.01i\\	0&0.99
		\end{matrix}\right)$&\\
		\hline
	\end{tabular}
	
	\begin{tabular}{|c|c|c|c|c|c|}
		\hline
		\diagbox{\small{a}}{$|\mathring{z}_{a,b}\rangle$}{\small{b}}&16&17&18&19&20\\
		\hline
		16&\diagbox{}{}&\diagbox{}{}&\diagbox{}{}&\diagbox{}{}&(1,-1.7 - 0.68i)\\
		\hline
		17&(1,0.87 - 0.48i)&\diagbox{}{}&(1,-0.82 + 0.58i)&(1,-0.76 + 0.75i)&\diagbox{}{}\\
		\hline
		18&(1,-1)&\diagbox{}{}&\diagbox{}{}&(1,1.21 + 0.14i)&\diagbox{}{}\\
		\hline
		19&(1,1.51 + 0.42i)&\diagbox{}{}&\diagbox{}{}&\diagbox{}{}&\diagbox{}{}\\
		\hline
		20&\diagbox{}{}&(1,0.88 - 0.59i)&(1,-1.20 - 0.13i)&(1,2.54 + 0.65i)&\diagbox{}{}\\
		\hline
	\end{tabular}
	
	}
    \end{table}
    
    \begin{table}[H]
	\scalebox{0.7}
	{
	\begin{tabular}{|c|c|c|c|}
		\hline
		\small{a}&21&22&23\\
		\hline
		$\mathring{g}_{a}$ &$\left(\begin{matrix}
		0.87&-0.06+0.086i\\	-0.06-0.085i&1.16
		\end{matrix}\right)$&$\left(\begin{matrix}
		0.97 &-0.13-0.45i\\	0&1.03
		\end{matrix}\right)$&$\left(\begin{matrix}
		0.98&-0.016+0.023i\\	0&1.02
		\end{matrix}\right)$\\
		\hline
		\small{a}&24&25&\\
		\hline
		$\mathring{g}_{a}$ &$\left(\begin{matrix}
		1.64&-0.17+0.24i\\	-0.05-0.07i&0.62
		\end{matrix}\right)$&$\left(\begin{matrix}
		1.04&-0.14-0.01i\\	0.26&0.99-0.003i
		\end{matrix}\right)$&\\
		\hline
	\end{tabular}
	
	\begin{tabular}{|c|c|c|c|c|}
		\hline
		\diagbox{\small{a}}{$|\mathring{z}_{a,b}\rangle$}{\small{b}}&21&23&24&25\\
		\hline
		22&(1,0.18 - 1.43i)&(1,-0.15 + 0.78i)&(1,0.078 - 0.58i)&(1,0.64 - 0.88i)\\
		\hline
		23&(1,0.18 + 0.26i)&\diagbox{}{}&(1,4.6 + 6.5i)&(1,-1.11 - 0.072i)\\
		\hline
		24&(1,5.72 + 8.08i)&\diagbox{}{}&\diagbox{}{}&(1,4.58 + 3.90i)\\
		\hline
		25&(1,-1.41 - 0.31i)&\diagbox{}{}&\diagbox{}{}&\diagbox{}{}\\
		\hline
	\end{tabular}
	}
    \end{table}

The flat geometry on $\sig_{\text{1-5}}$ is not unique. The position of $P_6$ can move continuously in $\R^4$ to lead to the continuous family of flat geometries on $\sig_{\text{1-5}}$. The continuous family of flat geometries result in the continuous family of real critical points. It implies that all these real critical points lead to degenerate Hessian matrices, in contrast to $A(\Delta_3)$ where the real critical point is nondegenerate. Therefore we develop the following additional procedure to generalize the analysis from $\Delta_3$ to $\sig_{\text{1-5}}$.

We label boundary spins $j_{mnk}$ by a triple of points $m\neq n\neq k=1,2,\cdots,5$, and label the internal spins $j_{mn6}$ by $m,n=1,2,\cdots,5$ and point $6$. The dual faces and spins are labelled in the dual cable diagram Fig.\ref{4complex}(b). We pick up 5 internal spins ${j}_{126},\,{j}_{136},\,{j}_{146},\,{j}_{156},\,{j}_{236}$ and their corresponding integrals in $A(\sigma_{\text{1-5}})$. The integrand is denoted by $\cz_{\sigma_{\text{1-5}}}$. Namely
\be
A(\sigma_{\text{1-5}})&=&\int_{\R^5} \rmd j_{126}\rmd j_{136}\rmd j_{146}\rmd j_{156}\rmd j_{236}\, \cz_{\sigma_{\text{1-5}}}\lt(j_{126},j_{136},j_{146},j_{156},j_{236}\rt),\label{Aandczapp}\\
\cz_{\sigma_{\text{1-5}}}&=&\sum_{\{k_h\}}\int_{\R^5} \prod_{\bar{h}=1}^5\mathrm{d} j_{\bar h}\, \prod_{h=1}^{10} 2 \l\,\t_{[-\epsilon,\l j^{\rm max}+\epsilon]}(\l j_h)\int [\rmd g\rmd \mathbf{z}] e^{\lambda S^{(k)}},\label{czintegralapp}
\ee
where other five internal spins ${j}_{246},\,{j}_{256},\,{j}_{346},\,{j}_{356},\,{j}_{456}$ are denoted by ${j}_{\bar{h}}$ ($\bar{h}=1,2,\cdots,5$). At the real critical point constructed above, the 5 areas $\mathring{j}_{126},\,\mathring{j}_{136},\,\mathring{j}_{146},\,\mathring{j}_{156},\,\mathring{j}_{236}$ are determined by the internal segment-lengths $\mathring{l}_{m6}$ ($m=1,2,\cdots,5$) via the Heron's formula. We focus on a neighborhood of $({j}_{126},\,{j}_{136},\,{j}_{146},\,{j}_{156},\,{j}_{236})\in\R^5$ around $(\mathring{j}_{126},\,\mathring{j}_{136},\,\mathring{j}_{146},\,\mathring{j}_{156},\,\mathring{j}_{236})$ such that the five $j$'s in the neighborhood uniquely correspond to the five segment-lengths ${l}_{m6}$, $m=2,\cdots,5$.

We generalize the analysis of $A(\Delta_3)$ to $\cz_{\sigma_{\text{1-5}}}$. $\cz_{\sigma_{\text{1-5}}}$ in Eq.(\ref{czintegralapp}) contain integrals with the external parameters 
\be
r=\lbrace {j}_{126},\,{j}_{136},\,{j}_{146},\,{j}_{156},\,{j}_{236},\, {j}_b,\, {\xi}_{eb}\rbrace
\ee 
which including not only boundary data but also 5 internal $j$'s. We focus on the integral in $\cz_{\sigma_{\text{1-5}}}$ at $k_h=0$. Given $r=\mathring{r}=\lbrace \mathring{j}_{126},\,\mathring{j}_{136},\,\mathring{j}_{146},\,\mathring{j}_{156},\,\mathring{j}_{236},\, \mathring{j}_b,\, \mathring{\xi}_{eb}\rbrace$, the integral has the real critical point $\lbrace \mathring{j}_{\bar{h}},\mathring{g}_a,\mathring{\textbf{z}}_{a,b}\rbrace$ corresponding to the flat geometry $\mathbf{g}(\mathring{r})$. The data of $\mathring{r}$ and the real critical point are given in the above tables. The Hessian matrix at $\mathring{x}$ is nondegenerate in $\cz_{\sigma_{\text{1-5}}}$.

The similar parametrizations as Eq.(\ref{ga1}), (\ref{ga2}), (\ref{zab}), and (\ref{jab}) for $g_a,\textbf{z}_{a,b},j_{\bar{h}}$ define the local coordinates $x\in \mathbb{R}^{195}$ covering a neighborhood $K$ of $\mathring{x}=(0,0,\cdots,0)$. We again express the spinfoam action as $S(r,x)$. The integral in $\cz_{\sigma_{\text{1-5}}}$ is of the same type as \eqref{typeint}} with $N=195$.

\subsection{Geometrical variations}

All the above data relate to the real critical point and the flat geometry with 10 deficit angles all zero. To give the curved geometries, we fix the boundary data $\mathring{j}_b, \mathring{\xi}_{eb}$ and deform the 5 internal segment-lengths $l_{m6}=\mathring{l}_{m6}+\delta l_{m6}$, $m=1,\cdots,5$. We randomly sample $\delta l_{m6}$ in the range $10^{-15}$ to $10^{-5}$. Each time, for the each new internal segment-lengths $l_{m6}$, we can solve for the coordinate $P_{6}$. Then, we repeat the procedure above to reconstruct the geometry and compute all the geometric quantities of triangulation: e.g. the areas, the 4-d normals of each tetrahedron, and the deficit angles. Some data of the deformation $\delta l_{m6}=(\delta l_{16},\delta l_{26},\delta l_{36},\delta l_{46},\delta l_{56})$ and the corresponding deficit angles $\delta_h$ are shown in Tables \ref{tab:delta101} and \ref{tab:delta102}, 
\begin{table}[H]
   \centering\caption{Deficit angles as $\delta l_{m6}=(3.0\times10^{-6},3.7\times10^{-6},-3.1\times10^{-6},-2.8\times10^{-6},-3.6\times10^{-6})$}
     \scalebox{0.9}
     {
    \begin{tabular}{|c|c|c|c|c|c|c|c|c|c|c|}
    \hline
    $\delta_1$ & $\delta_2$&$\delta_3$&$\delta_4$&$\delta_5$&$\delta_6$&$\delta_7$&$\delta_8$&$\delta_9$&$\delta_{10}$&$\delta$\\
    \hline
    $6.1\times10^{-5}$&$2.6\times10^{-4}$&$1.1\times10^{-4}$&$1.4\times10^{-4}$&$4.6\times10^{-5}$&$1.4\times10^{-5}$&$1.8\times10^{-5}$&$1.3\times10^{-4}$&$1.1\times10^{-4}$&$4.1\times10^{-5}$&$1.2\times10^{-4}$\\
    \hline
    \end{tabular}
    \label{tab:delta101}
    }
    \centering\caption{Deficit angles as $\delta l_{m6}=(-3.\times10^{-8},5.0\times10^{-8},3.4\times10^{-8},3.1\times10^{-8},4.0\times10^{-8})$}
    \scalebox{0.9}
     {
    \begin{tabular}{|c|c|c|c|c|c|c|c|c|c|c|}
    \hline
    $\delta_1$ & $\delta_2$&$\delta_3$&$\delta_4$&$\delta_5$&$\delta_6$&$\delta_7$&$\delta_8$&$\delta_9$&$\delta_{10}$&$\delta$\\
    \hline
   $1.5\times10^{-6}$& $6.4\times10^{-6}$& $2.8\times10^{-6}$&    $3.5\times10^{-6}$&$ 1.1\times10^{-6}$& $3.6\times10^{-7}$&
    $4.5\times10^{-7}$& $3.3\times10^{-6}$& $2.8\times10^{-6}$& $1.0\times10^{-6}$&$2.9\times10^{-6}$\\
    \hline
    \end{tabular}
    }
    \label{tab:delta102}
\end{table}
Here $\delta$ is the average deficit angle $\delta=\sqrt{\frac{1}{10}\sum_{h=1}^{10}\delta^2_h}$.

Fixing $\mathring{j}_b, \mathring{\xi}_{eb}$, varying $l_{m6}=\mathring{l}_{m6}+\delta l_{m6}$ results in varying the 5 areas in $r$ e.g. $j_{126}=\mathring{j}_{126}+\delta j_{126}, j_{136}=\mathring{j}_{136}+\delta j_{136}, \cdots$. Thus we obtain the deformation of external data $r=\mathring{r}+\delta r$ of $\cz_{\sigma_{\text{1-5}}}$. We denote by $r_l$ the external data obtained by sampling $\delta l_{m6}$, and denote the Regge geometries by $\mathbf{g}(r_l)$. There are 4 degrees of freedom of $\delta l_{m6}$ still resulting in flat geometries, whereas there is 1 degree of freedom of $\delta l_{m6}$ resulting in curved geometries.

\subsection{Complex critical points and numerical results} 

We apply the Newton-like recursive method similar to Section \ref{Numerical solving complex critical points} to numerically compute complex critical points $Z(r_l)$ for all $r_l$. The absolute errors in the case $\gamma=1, n=3$ for some average deficit angles are shown in Table \ref{error1-5}.
\begin{table}[H]
    \centering
    \begin{tabular}{|c|c|c|c|c|c|c|c|c|}
        \hline
        $\delta$ & $1.2\times10^{-4}$&$1.2\times10^{-5}$& $2.1\times10^{-6}$&$6.5\times10^{-7}$&$1.3\times10^{-8}$&$1.2\times10^{-10}$&$1.5\times10^{-11}$&$1.4\times10^{-12}$\\
        \hline
        $\varepsilon$ & $4.0\times10^{-15}$&$2.1\times10^{-19}$&$2.0\times10^{-22}$&$2.0\times10^{-27}$&$2.3\times10^{-31}$&$2.3\times10^{-39}$&$5.0\times10^{-43}$&$5.0\times10^{-47}$ \\
        \hline
    \end{tabular}
    \label{error1-5}
\end{table}
$Z(r_l)$ is still in the real plane if $r_l$ corresponds to the flat geometry, whereas $Z(r_l)$ is away from the real plane if $r_l$ corresponds to the curved geometry.

Once we have complex critical points $Z(r_l)$ for the curved geometries $\textbf{g}(r_l)$, we numerically compute the analytic continued action $\cs(r_l, Z(r_l))$ at complex critical points and the difference $\delta \ci(r_l)=\cs(r_l, Z(r_l))-S(r_l,x_0)$ where $x_0$ is the pseudo-critical point of $S(r_l,x)$ (recall Section \ref{Numerical solving complex critical points}). We have $S(r_l,x_0)=-i\ci_{R}[\mathbf{g}(r)]+i\varphi$, {where $\varphi$ only relates to the boundary data and is independent of $l_{m6}$ as confirmed by numerical tests (see also \cite{Han:2011re} for the analytic argument)}. Some numerical results of $\delta\ci$ are shown in Table \ref{deltaI1-5}

\begin{table}[H]
    \centering
    \scalebox{0.8}
    {
    \begin{tabular}{|c|c|c|c|c|c|}
        \hline
        $\delta$ & $1.2\times10^{-4}$& $2.1\times10^{-6}$&$3.8\times10^{-8}$&$6.5\times10^{-10}$&$6.5\times10^{-12}$\\
        \hline
        $\delta\ci$ & $-1.2\times10^{-12}+4.5\times10^{-10}i$&$-3.8\times10^{-16}+1.4\times10^{-13}i$&$-1.3\times10^{-19}+4.7\times10^{-17}i$&$-3.8\times10^{-23}+1.4\times10^{-20}i$&$-3.8\times10^{-27}+1.4\times10^{-24}i$ \\
        \hline
    \end{tabular}
    }
    \label{deltaI1-5}
\end{table}
The best-fit function is $\delta\ci=-a_2(\gamma)\delta^2+O(\delta^3)$, the best fit coefficient and the corresponding fitting errors at $\g=1$ is:
\be
a_2=8.88\times 10^{-5}_{\pm 10^{-12}} - i 0.033_{\pm 10^{-10}}.
\ee
We use FIG.\ref{thetaS}(b) to demonstrate the excellent agreement between the numerical data and the best-fit function. 

As a result, we obtain the following large-$\lambda$ contribution to $\cz_{\sig_{\text{1-5}}}$ and $A({\sig_{\text{1-5}}})$ from the neighborhood around $(\mathring{r},\mathring{x})$ 
\be
\cz_{\sig_{\text{1-5}}}&\sim&\lt(\frac{1}{\l}\rt)^{\frac{155}{2}} e^{i\l \varphi}\sn'_{l}\, e^{-i\l \ci_{R}[\textbf{g}(r_l)]-\l a_2(\gamma)\delta(r_l)^2+O(\delta^3)}\lt[1+O(1/\l)\rt],\\
A({\sig_{\text{1-5}}})&\sim&\lt(\frac{1}{\l}\rt)^{\frac{155}{2}}e^{i\l \varphi}\int \prod_{m=1}^5\rmd l_{m6}\sn_{l}\, e^{-i\l \ci_{R}[\textbf{g}(r_l)]-\l a_2(\gamma)\delta(r_l)^2+O(\delta^3)}\lt[1+O(1/\l)\rt],
\ee 
where we have made the local changes of variables from ${j}_{126},\,{j}_{136},\,{j}_{146},\,{j}_{156},\,{j}_{236}$ to $l_{m6}$, and the {Jacobian} $\cj_l=|\det(\partial j/\partial l)|$ is absorbed in $\sn_l=\cj_l\sn_l'$. The spinfoam amplitude $A({\sig_{\text{1-5}}})$ reduces to the integral over geometries $\mathbf{g}(r_l)$ in the semiclassical regime.

The Jacobian $\cj_l$ reads:
\small
\be
&&\frac{l_{16} l_{26} l_{36} l_{46} l_{56}\big(l_{14}^2+l_{16}^2-l_{46}^2\big) \big(l_{15}^2+l_{16}^2-l_{56}^2\big)\lt\{\left[\left(l_{16}^2-l_{36}^2\right) \left(l_{26}^2-l_{36}^2\right)-l_{13}^2 l_{23}^2\right] l_{12}^2+\left(l_{16}^2-l_{26}^2\right) \left[\left(l_{36}^2-l_{26}^2\right) l_{13}^2+l_{23}^2 \left(l_{16}^2-l_{36}^2\right)\right]\rt\}}{16\sqrt{-l_{12}^4+2 \left(l_{16}^2+l_{26}^2\right) l_{12}^2-\left(l_{16}^2-l_{26}^2\right){}^2} \sqrt{-l_{13}^4+2 \left(l_{16}^2+l_{36}^2\right) l_{13}^2-\left(l_{16}^2-l_{36}^2\right){}^2}\sqrt{-l_{23}^4+2 \left(l_{26}^2+l_{36}^2\right) l_{23}^2-\left(l_{26}^2-l_{36}^2\right){}^2}}\nonumber\\
&&\frac{1}{\sqrt{-l_{14}^4+2 \left(l_{16}^2+l_{46}^2\right) l_{14}^2-\left(l_{16}^2-l_{46}^2\right){}^2} \sqrt{-l_{15}^4+2 \left(l_{16}^2+l_{56}^2\right) l_{15}^2-\left(l_{16}^2-l_{56}^2\right){}^2}}
\nonumber.
\ee


\bibliographystyle{jhep}
\bibliography{ref}

\providecommand{\href}[2]{#2}\begingroup\raggedright\begin{thebibliography}{10}

\bibitem{Thiemann:2007pyv}
T.~Thiemann, {\em {Modern Canonical Quantum General Relativity}}.
\newblock Cambridge Monographs on Mathematical Physics. Cambridge University
  Press, 2007.

\bibitem{Rovelli:2014ssa}
C.~Rovelli and F.~Vidotto, {\em {Covariant Loop Quantum Gravity}: {An
  Elementary Introduction to Quantum Gravity and Spinfoam Theory}}.
\newblock Cambridge Monographs on Mathematical Physics. Cambridge University
  Press, 11, 2014.

\bibitem{Perez2012}
A.~Perez, {\it {The Spin Foam Approach to Quantum Gravity}},  {\em Living
  Rev.Rel.} {\bf 16} (2013) 3, [\href{http://arxiv.org/abs/1205.2019}{{\tt
  arXiv:1205.2019}}].

\bibitem{Rovelli:2010bf}
C.~Rovelli, {\it {Loop quantum gravity: the first twenty five years}},  {\em
  Class. Quant. Grav.} {\bf 28} (2011) 153002,
  [\href{http://arxiv.org/abs/1012.4707}{{\tt arXiv:1012.4707}}].

\bibitem{Ashtekar:2017yom}
A.~Ashtekar and J.~Pullin, eds., {\em {Loop Quantum Gravity}: {The First 30
  Years}}, vol.~4 of {\em 100 Years of General Relativity}.
\newblock World Scientific, 2017.

\bibitem{Ashtekar:2021kfp}
A.~Ashtekar and E.~Bianchi, {\it {A short review of loop quantum gravity}},
  {\em Rept. Prog. Phys.} {\bf 84} (2021), no.~4 042001,
  [\href{http://arxiv.org/abs/2104.04394}{{\tt arXiv:2104.04394}}].

\bibitem{Engle:2007uq}
J.~Engle, R.~Pereira, and C.~Rovelli, {\it {The Loop-quantum-gravity
  vertex-amplitude}},  {\em Phys. Rev. Lett.} {\bf 99} (2007) 161301,
  [\href{http://arxiv.org/abs/0705.2388}{{\tt arXiv:0705.2388}}].

\bibitem{Engle:2007wy}
J.~Engle, E.~Livine, R.~Pereira, and C.~Rovelli, {\it {LQG vertex with finite
  Immirzi parameter}},  {\em Nucl. Phys. B} {\bf 799} (2008) 136--149,
  [\href{http://arxiv.org/abs/0711.0146}{{\tt arXiv:0711.0146}}].

\bibitem{Rovelli:2005yj}
C.~Rovelli, {\it {Graviton propagator from background-independent quantum
  gravity}},  {\em Phys. Rev. Lett.} {\bf 97} (2006) 151301,
  [\href{http://arxiv.org/abs/gr-qc/0508124}{{\tt gr-qc/0508124}}].

\bibitem{Livine:2007vk}
E.~R. Livine and S.~Speziale, {\it {A New spinfoam vertex for quantum
  gravity}},  {\em Phys. Rev. D} {\bf 76} (2007) 084028,
  [\href{http://arxiv.org/abs/0705.0674}{{\tt arXiv:0705.0674}}].

\bibitem{Freidel:2007py}
L.~Freidel and K.~Krasnov, {\it {A New Spin Foam Model for 4d Gravity}},  {\em
  Class. Quant. Grav.} {\bf 25} (2008) 125018,
  [\href{http://arxiv.org/abs/0708.1595}{{\tt arXiv:0708.1595}}].

\bibitem{Conrady:2008mk}
F.~Conrady and L.~Freidel, {\it {On the semiclassical limit of 4d spin foam
  models}},  {\em Phys. Rev.} {\bf D78} (2008) 104023,
  [\href{http://arxiv.org/abs/0809.2280}{{\tt arXiv:0809.2280}}].

\bibitem{Barrett:2009mw}
J.~W. Barrett, R.~J. Dowdall, W.~J. Fairbairn, F.~Hellmann, and R.~Pereira,
  {\it {Lorentzian spin foam amplitudes: Graphical calculus and asymptotics}},
  {\em Class. Quant. Grav.} {\bf 27} (2010) 165009,
  [\href{http://arxiv.org/abs/0907.2440}{{\tt arXiv:0907.2440}}].

\bibitem{Han:2011re}
M.~Han and M.~Zhang, {\it {Asymptotics of Spinfoam Amplitude on Simplicial
  Manifold: Lorentzian Theory}},  {\em Class. Quant. Grav.} {\bf 30} (2013)
  165012, [\href{http://arxiv.org/abs/1109.0499}{{\tt arXiv:1109.0499}}].

\bibitem{Han:2013gna}
M.~Han and T.~Krajewski, {\it {Path Integral Representation of Lorentzian
  Spinfoam Model, Asymptotics, and Simplicial Geometries}},  {\em Class. Quant.
  Grav.} {\bf 31} (2014) 015009, [\href{http://arxiv.org/abs/1304.5626}{{\tt
  arXiv:1304.5626}}].

\bibitem{Kaminski:2017eew}
W.~Kaminski, M.~Kisielowski, and H.~Sahlmann, {\it {Asymptotic analysis of the
  EPRL model with timelike tetrahedra}},  {\em Class. Quant. Grav.} {\bf 35}
  (2018), no.~13 135012, [\href{http://arxiv.org/abs/1705.02862}{{\tt
  arXiv:1705.02862}}].

\bibitem{Liu:2018gfc}
H.~Liu and M.~Han, {\it {Asymptotic analysis of spin foam amplitude with
  timelike triangles}},  {\em Phys. Rev. D} {\bf 99} (2019), no.~8 084040,
  [\href{http://arxiv.org/abs/1810.09042}{{\tt arXiv:1810.09042}}].

\bibitem{Simao:2021qno}
J.~D. Sim\~ao and S.~Steinhaus, {\it {Asymptotic analysis of spin-foams with
  time-like faces in a new parameterisation}},
  \href{http://arxiv.org/abs/2106.15635}{{\tt arXiv:2106.15635}}.

\bibitem{Dona:2020yao}
P.~Dona and S.~Speziale, {\it {Asymptotics of lowest unitary SL(2,C) invariants
  on graphs}},  {\em Phys. Rev. D} {\bf 102} (2020), no.~8 086016,
  [\href{http://arxiv.org/abs/2007.09089}{{\tt arXiv:2007.09089}}].

\bibitem{Engle:2020ffj}
J.~S. Engle, W.~Kaminski, and J.~R. Oliveira, {\it {Addendum to
  \textquoteleft{}EPRL/FK asymptotics and the flatness
  problem\textquoteright{}}},  \href{http://arxiv.org/abs/2012.14822}{{\tt
  arXiv:2012.14822}}. [Addendum: Class.Quant.Grav. 38, 119401 (2021)].

\bibitem{Hellmann:2012kz}
F.~Hellmann and W.~Kaminski, {\it {Geometric asymptotics for spin foam lattice
  gauge gravity on arbitrary triangulations}},
  \href{http://arxiv.org/abs/1210.5276}{{\tt arXiv:1210.5276}}.

\bibitem{Bonzom:2009hw}
V.~Bonzom, {\it {Spin foam models for quantum gravity from lattice path
  integrals}},  {\em Phys. Rev. D} {\bf 80} (2009) 064028,
  [\href{http://arxiv.org/abs/0905.1501}{{\tt arXiv:0905.1501}}].

\bibitem{Han:2013hna}
M.~Han, {\it {On Spinfoam Models in Large Spin Regime}},  {\em Class. Quant.
  Grav.} {\bf 31} (2014) 015004, [\href{http://arxiv.org/abs/1304.5627}{{\tt
  arXiv:1304.5627}}].

\bibitem{Gozzini:2021kbt}
F.~Gozzini, {\it {A high-performance code for EPRL spin foam amplitudes}},
  \href{http://arxiv.org/abs/2107.13952}{{\tt arXiv:2107.13952}}.

\bibitem{Engle:2011pmf}
J.~Engle, {\it {The Plebanski sectors of the EPRL vertex}},  {\em Class. Quant.
  Grav.} {\bf 28} (2011) 225003, [\href{http://arxiv.org/abs/1301.2214}{{\tt
  arXiv:1301.2214}}]. [Erratum: Class.Quant.Grav. 30, 049501 (2013)].

\bibitem{Rovelli1995}
C.~Rovelli and L.~Smolin, {\it {Discreteness of area and volume in quantum
  gravity}},  {\em Nuclear Physics B} {\bf 442} (May, 1995) 593--619.

\bibitem{ALarea}
A.~Ashtekar and J.~Lewandowski, {\it {Quantum theory of geometry. 1: Area
  operators}},  {\em Class.Quant.Grav.} {\bf 14} (1997) A55--A82,
  [\href{http://arxiv.org/abs/gr-qc/9602046}{{\tt gr-qc/9602046}}].

\bibitem{10.1007/BFb0074195}
A.~Melin and J.~Sj{\"o}strand, {\it Fourier integral operators with
  complex-valued phase functions},  in {\em Fourier Integral Operators and
  Partial Differential Equations} (J.~Chazarain, ed.), (Berlin, Heidelberg),
  pp.~120--223, Springer Berlin Heidelberg, 1975.

\bibitem{Hormander}
L.~Hormander, {\em The Analysis of Linear Partial Differential Operators I}.
\newblock Springer-Verlag Berlin, 1983.

\bibitem{Dona:2020tvv}
P.~Dona, F.~Gozzini, and G.~Sarno, {\it {Numerical analysis of spin foam
  dynamics and the flatness problem}},
  \href{http://arxiv.org/abs/2004.12911}{{\tt arXiv:2004.12911}}.

\bibitem{Asante:2020qpa}
S.~K. Asante, B.~Dittrich, and H.~M. Haggard, {\it {Effective Spin Foam Models
  for Four-Dimensional Quantum Gravity}},  {\em Phys. Rev. Lett.} {\bf 125}
  (2020), no.~23 231301, [\href{http://arxiv.org/abs/2004.07013}{{\tt
  arXiv:2004.07013}}].

\bibitem{Bahr:2016hwc}
B.~Bahr and S.~Steinhaus, {\it {Numerical evidence for a phase transition in 4d
  spin foam quantum gravity}},  {\em Phys. Rev. Lett.} {\bf 117} (2016), no.~14
  141302, [\href{http://arxiv.org/abs/1605.07649}{{\tt arXiv:1605.07649}}].

\bibitem{Delcamp:2016dqo}
C.~Delcamp and B.~Dittrich, {\it {Towards a phase diagram for spin foams}},
  {\em Class. Quant. Grav.} {\bf 34} (2017), no.~22 225006,
  [\href{http://arxiv.org/abs/1612.04506}{{\tt arXiv:1612.04506}}].

\bibitem{Banburski:2014cwa}
A.~Banburski, L.-Q. Chen, L.~Freidel, and J.~Hnybida, {\it {Pachner moves in a
  4d Riemannian holomorphic Spin Foam model}},  {\em Phys. Rev. D} {\bf 92}
  (2015), no.~12 124014, [\href{http://arxiv.org/abs/1412.8247}{{\tt
  arXiv:1412.8247}}].

\bibitem{Han:2020fil}
M.~Han, Z.~Huang, H.~Liu, and D.~Qu, {\it {Numerical computations of
  next-to-leading order corrections in spinfoam large-$j$ asymptotics}},  {\em
  Phys. Rev. D} {\bf 102} (2020), no.~12 124010,
  [\href{http://arxiv.org/abs/2007.01998}{{\tt arXiv:2007.01998}}].

\bibitem{Dona:2019dkf}
P.~Dona, M.~Fanizza, G.~Sarno, and S.~Speziale, {\it {Numerical study of the
  Lorentzian Engle-Pereira-Rovelli-Livine spin foam amplitude}},  {\em Phys.
  Rev.} {\bf D100} (2019), no.~10 106003,
  [\href{http://arxiv.org/abs/1903.12624}{{\tt arXiv:1903.12624}}].

\bibitem{qudx_Delta3.org}
D.~Qu.
  \url{https://github.com/dqu2017/Complex-critical-points-and-curved-geometries-in-spinfoam-quantum-gravity},
  2021.

\bibitem{Han:2021rjo}
M.~Han and H.~Liu, {\it {Analytic Continuation of Spin foam Models}},  4, 2021.
\newblock \href{http://arxiv.org/abs/2104.06902}{{\tt arXiv:2104.06902}}.

\end{thebibliography}\endgroup

\end{document}